\documentclass[%
 reprint,
superscriptaddress,
 amsmath,amssymb,
 aps,
physrev,
floatfix,
showkeys
]{revtex4-2}

\usepackage{graphicx}
\usepackage{xcolor}
\DeclareMathOperator*{\argmax}{argmax}

\begin{document}
\title{Method for SOFI-based spatial super-resolution in nanosensing with blinking emitters}
\author{Alexander Mikhalychev}
\email{alexander.mikhalychev@atomicus-software.com}
\affiliation{Atomicus Sp. z o.o., Ul. Aleja Grunwaldzka 163,
Gdansk, 80-266 Poland}
\affiliation{Atomicus GmbH, Amalienbadstr. 41C, Karlsruhe, 76227 Germany}
\author{Alex Ulyanenkov}
\affiliation{Atomicus GmbH, Amalienbadstr. 41C, Karlsruhe, 76227 Germany}

\date{\today}


\begin{abstract}
We propose a method of spatial resolution enhancement in metrology (thermometry, magnetometry, pH estimation, and similar methods) with blinking fluorescent nanosensors by combining sensing with super-resolution optical fluctuation imaging (SOFI). {\color{black} By utilizing the idea of quantum super-resolution imaging by photon statistics (QSIPS), the applicability of the proposed methodology is extended to low-brightness regime. Efficiency of the approach is demonstrated by numerical simulations performed for several model configurations, representing step-like and continuous variation of the sensed parameter, high and low brightness regimes, 1- and 2-dimensional structures.} The 2nd and 4th order cumulant images provide improvement of the contrast and enable successful reconstruction of smaller features of the modeled parameter distribution relatively to the intensity-based approach. We believe that blinking fluorescent sensing agents being complemented with the developed image analysis technique could be utilized routinely in the life science sector for recognizing the local changes in the spectral response of blinking fluorophores, e.g. delivered targetly to the specific cell or even organelle. It is extremely useful for the local measurements of living cells' physical parameters changes due to applying any external ``forces'', including disease effect, aging, healing, or response to the treatment. 
\end{abstract}

\maketitle

\section{Introduction}

Fluorescent nanosensors represent a powerful tool for measurement of temperature, magnetic field, pH, and other parameters with high accuracy at small scales \cite{zhou2020advances,norman2021novel,wu2022recent,radtke2019nanoscale,xu2023recent,baranov2011silicon,anisimov2016optical}. In particular, the dependence of their fluorescent signal on the mentioned parameters can be utilized for investigation of processes in living cells \cite{wu2022recent,fu2007characterization,hui2010two}. To ensure high spatial resolution, one needs to localize the nanoemitters precisely and to distinguish signals from closely located nanosensors. This requirement can be fulfilled, for example, by placement of fluorescent nanodiamonds or quantum dots at a manipulator needle \cite{blakley2015room,blakley2016fiber,chang2017nanoscale,pelliccione2016scanned,malykhin2022control}. This approach requires precise mechanical control and does not allow to analyze environment parameters at several points of the sample simultaneously. For a large number of nanosensors, simultaneously attached to the sample, one can apply deterministic super-resolution approaches based on selective (localized) excitation of emitters (stimulated emission depletion (STED), ground state depletion (GSD), reversible saturable optical fluorescence transitions between spin levels (spin-RESOLFT), structured illumination imaging) \cite{hell1994breaking,rittweger2009far,rittweger2009sted,maurer2010far,classen2017superresolution}, as well as stochastic methods (stochastic optical reconstruction microscopy (STORM), photoactivated localization microscopy (PALM), points accumulation for imaging in nanoscale topography (PAINT)) \cite{hess2006ultra,rust2006sub,schnitzbauer2017super,gu2013super}, ensuring independent localization of single emitters due to their blinking. Those approaches are also quite sophisticated and require the nanosensors to remain in the inactive state most of the time, thus increasing needed exposition time and potential photo-toxicity.

Alternatively to the super-resolution approaches, based on scanning or selective excitation (activation) of emitters, wide-field microscopy methods exploiting specific temporal statistics of fluorescent objects, i.e. stochastic blinking (in super-resolution optical fluctuation imaging (SOFI) \cite{dertinger2009fast,dertinger2010achieving,dertinger2013advances} and quantum super-resolution imaging by
photon statistics (QSIPS) \cite{picariello2025imaging}) or photon antibunching \cite{schwartz2012improved,schwartz2013superresolution,monticone2014beating} have been successfully implemented. The idea is based on postprocessing of time-dependent fluorescent signal: its efficient exponentiation by considering correlations instead of raw intensity, elimination of cross-terms from different emitters, and subsequent enhancement of spatial resolution. Analysis of the fluorescence blinking does not preclude usage of the ``degrees of freedom'' suitable for the environment sensing, such as the dependence of the spectrum or fluorescence lifetime on the temperature, acidity, pressure, etc. In the current contribution we suggest to combine sensing capabilities of nanoparticles with SOFI-based super-resolution {\color{black}(including QSIPS for low-brightness emitters)} for more accurate localization of the signal source. Is is important to note that the information, used by SOFI to achieve sub-Rayleigh spatial resolution, is already present in the collected fluorescent signal and does not introduce any additional complications to the measurement.

\begin{figure*}[tp]
    \centering
    \includegraphics[width=0.9\linewidth]{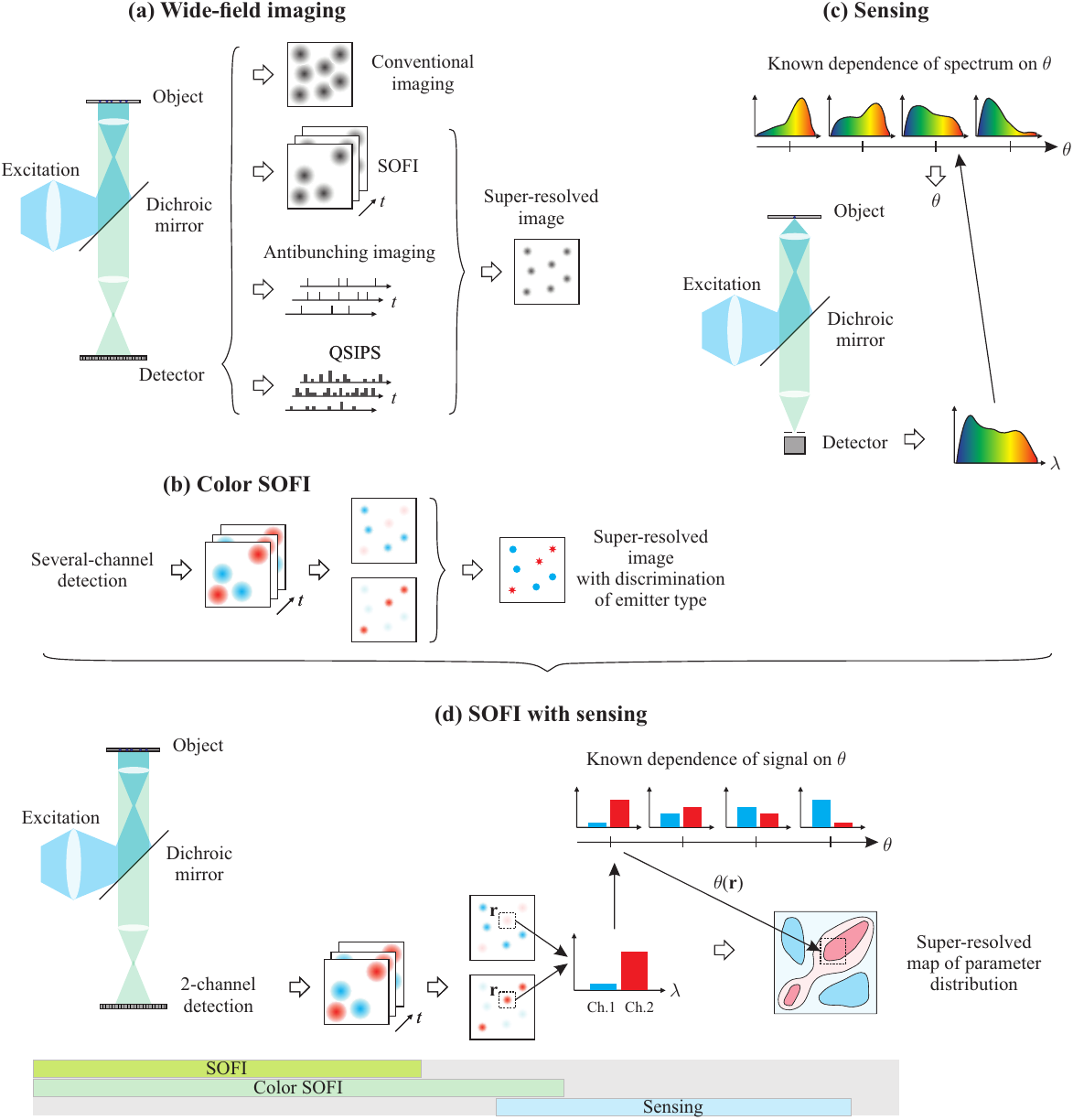}
    \caption{Techniques of imaging and sensing with fluorescent emitters, related to the proposed approach. (a) Wide-field imaging: conventional intensity accumulation and super-resolving approaches based on analysis of the temporal statistics (SOFI \cite{dertinger2009fast,dertinger2010achieving,dertinger2013advances}, antibunching imaging \cite{schwartz2012improved,schwartz2013superresolution,monticone2014beating}, and QSIPS \cite{picariello2025imaging}). (b) Color SOFI \cite{gallina2013resolving,grussmayer2020spectral,liu2020narrow,glogger2021multi,Chizhik2016CND}: localization of emitters with discrimination of their types. (c) Sensing: estimation of the continuous (real-valued) parameter $\theta$ from the signal spectrum. (d) Combining SOFI with sensing: estimation of distribution of the continuous (real-valued) parameter $\theta(\mathbf{r})$ from the spatially resolved signal with two spectral channels. Horizontal bars below the scheme show the relation of the proposed method steps with other imaging and sensing modalities. Similarly to color SOFI, the cumulant images are constructed for the two channels separately. Then, for each pixel, the pairs of values are analyzed in the spirit of sensing to get the local parameter estimate $\theta(\mathbf{r})$. The method results in construction of a map for the spatial distribution of the parameter. For clarity, spectral sensing is assumed in panels (c) and (d). }
    \label{fig:methods}
\end{figure*}

The place of the developed SOFI-based sensing approach among imaging and sensing techniques is schematically shown in Fig.~\ref{fig:methods}. Wide-field imaging is aimed at \textit{localization} of fluorescent emitters by projecting their signal on a camera (matrix detector) --- Fig.~\ref{fig:methods}(a). Basically, the intensity is accumulated without additional spectral or temporal splitting of the signal into multiple channels. Pixel-wise analysis of temporal statistics (frames in SOFI, numbers of photons in QSIPS and antibunching imaging) enables construction of a sharper image with spatial super-resolution. The so called ``color SOFI'' \cite{gallina2013resolving,grussmayer2020spectral,liu2020narrow,glogger2021multi,Chizhik2016CND} (Fig.~\ref{fig:methods}(b)) applies SOFI to several (typically two) spectral channels, thus allowing to \textit{localize} and \textit{classify} emitters (discriminate between several species). The primary task of sensing (Fig.~\ref{fig:methods}(c)) is to \textit{read-out a continuous} (real-valued) \textit{parameter} from a preliminary localized nanosensor. In contrast to imaging, sensing does not require a multi-pixel camera (a ``bucket'' photodetector is sufficient), but uses multichannel signal acquisition along another ``degree of freedom'' \cite{zhou2020advances,norman2021novel,wu2022recent,radtke2019nanoscale,xu2023recent,baranov2011silicon,anisimov2016optical}: wavelength, emission delay in pulsed excitation regime (fluorescence lifetime), signal dependence on alternating magnetic field frequency, etc. The proposed combination of super-resolved imaging with sensing (Fig.~\ref{fig:methods}(d)) ensures \textit{spatially resolved read-out of a continuous parameter} from a multichannel multi-pixel detected signal with collected temporal statistics.

Combining SOFI with sensing is not as trivial as it may seem at the first glance. First, we focus on the case when the parameter of interest may vary at the scales smaller that the classical resolution limit, determined by the point spread function (PSF) of the used microscope. Therefore, the assumptions of slowly varying parameters, used for example in balanced SOFI (bSOFI) \cite{geissbuehler2012mapping}, are not generally valid here. Second, SOFI imposes less restrictions on separate frames of the acquired timeseries in comparison with single-molecule localization techniques \cite{hell1994breaking,rittweger2009far,rittweger2009sted,maurer2010far}: it allows smaller numbers of detected photons per frame and higher probability of closely positioned emitters being active during the same frame. Thus, one cannot reliably perform reconstruction of the sensed environmental parameter for a single-frame dataset or for a separate localized emitter as it is done in spectrally resolved STORM, PAINT, and PALM techniques \cite{pfender2014single,zhang2015ultrahigh,mlodzianoski2016super,bongiovanni2016multi,moon2017spectrally}. Third, SOFI requires analysis of the whole acquired timeseries, relies on specific temporal fluctuations of the signal, and yields nonlinear dependence of the response on the emitters' brightness \cite{geissbuehler2012mapping,yi2019moments,pawlowska2021embracing}. Therefore, conversion of each separate frame into the investigated parameter map before processing of the entire timeseries would alter the fluctuations and preclude further application of SOFI for spatial resolution enhancement. On the other hand, application of SOFI to multichannel frame data (for example, spectra) would tremendously enhance the shot noise, while nonlinearity of the response would require additional treatment.

To resolve the outlined issues and combine SOFI with sensing, we propose to ``compress'' the sensing degree of freedom to just two channels, apply SOFI {\color{black} (or QSIPS)} to the timeseries corresponding to each of the two channels separately, and infer the sensed parameter map from the obtained cumulant images for those channels (Fig.~\ref{fig:methods}(d)). Linearity of the compression ensures preservation of the temporal fluctuations of the signal and applicability of SOFI to spatial resolution enhancement. The idea of using two channels for single-parameter inference is closely related to ratiometric imaging \cite{bruni2017two,klymchenko2022fluorescent}, where two spectral channels can be used for self-referencing \cite{schaferling2012art} to reduce the adverse effect of absolute intensity fluctuations \cite{bruni2017two,schaferling2012art,demchenko2009monitoring}. Nonlinearity of the resulting response of SOFI \cite{geissbuehler2012mapping,yi2019moments,pawlowska2021embracing} can be compensated analytically by inversion of the nonlinear relation between the sensed parameter of interest and cumulant values.

\color{black}
SOFI-based sensing relying on the signal ratio between two channels has recently been demonstrated in the form of FRET-SOFI \cite{valenta2025super}, where the donor-donor and donor-acceptor emission channels were used for estimation of F\"orster resonance energy transfer (FRET) efficiency. In that case, the two spectral channels were naturally determined by the two kinds of particles (donors and acceptors). In contrast, we consider sensing with just one kind of particles --- nanosensors with their (multichannel) spectral or temporal response depending on the parameter of interest. An important step of building the proposed sensing technique is definition of the two appropriate channels (``compression'' of the signal). The reconstruction of the sensed parameter by inverting its relation with ratio of the signals is another key feature, which distinguishes our approach. 
\color{black}

The idea of combined usage of blinking statistics of fluorophores for super-resolution and certain other degree of freedom for getting additional information is also close to color SOFI \cite{gallina2013resolving,grussmayer2020spectral,liu2020narrow,glogger2021multi,Chizhik2016CND} (Fig.~\ref{fig:methods}(b,d)). However, the tasks, solved by color SOFI and our approach, are different. Color SOFI deals with discrimination between several (two or three) discrete types of fluorescent emitters, serving as biological labels. In contrast, the method proposed in this paper, deals with continuous-variable sensing: the emitters are of the same type, but their fluorescent signal differs due to variations of local environment. The procedure of converting a two-channel super-resolved image (typical for color SOFI as well) into a map of the sensed parameter represents the essence of our approach and is novel relatively to other techniques.

To demonstrate principles of the proposed approach and show its feasibility, we applied it to multiple synthetic (simulated) datasets and quantified the performance and resolution enhancement. {\color{black} The presented simulations demonstrate combination of sensing with SOFI for bright nanosensors in 2-dimensional (square grid) and 1-dimensional (filaments) arrangements and with QSIPS for square grid of nanoemitters operating in a few-photon regime.} Information from both auto- and cross-cumulants \cite{dertinger2010achieving} with zero time lag was combined to reduce the influence of the shot noise and improve the accuracy of sensing. {\color{black} The improvement of the mean square error of the sensed parameter estimation reached the value $10^2$ for the 4th order cumulant relatively to traditional intensity-based sensing. }

The paper is organized as follows. Section \ref{sec:theory} recalls the basic ideas of SOFI, {\color{black}limitations of the technique, and its extension to low-brightness regime by QSIPS}. We present mathematical treatment of sensing as parameter estimation from multichannel signal, discuss optimal choice of the channels, and introduce the concept and the list of procedures constituting the proposed technique of spatially super-resolving sensing. Section~\ref{sec:modeling} is devoted to application of the approach to synthetic datasets and analysis of the resolution (accuracy of parameter estimation) improvement. Conclusions on the results and possible directions of further research are presented in Section~\ref{sec:conclusions}.

\section{Basic ideas}
\label{sec:theory}

\subsection{Super-resolution optical fluctuation imaging}

The commonly used model of wide-field fluorescence imaging includes a sample composed of $N$ incoherent florescent emitters, located at positions $\textbf{r}_k$ \cite{dertinger2009fast}. Fluctuations-based approaches assume that the emitters are independently blinking, stochastically switching between ``bright'' and ``dark'' states. The acquisition time is split into a finite number of frames. The fluorescence signal in the frame $t$ ($t = 1$, \ldots, $N_\text{f}$, where $N_\text{f}$ is the number of frames) can be described as \cite{dertinger2009fast}
\begin{equation}
    \label{eq:basic_signal}
    F(\mathbf{r}, t) = \sum_k U(\mathbf{r} - \mathbf{r}_k) \varepsilon_k s_k(t),
\end{equation}
where $U(\Delta \mathbf{r})$ is the point-spread function (PSF) of the optical system, $\varepsilon_k$ is the constant molecular brightness of the $k$-th emitter, and $s_k(t)$ is its time-dependent fluctuation ($s_k(t) = 1$ if the $k$-th emitter remains ``bright'' during the $t$-th frame; $s_k(t) = 0$ for the ``dark'' state). 

The idea of resolution enhancement in the wide-field microscopy (including quantum imaging, SOFI, antibunching imaging, and QSIPS) is to construct signal with power-law dependence on $U(\Delta \mathbf{r})$ and to take into account that, for a localized PSF, $U^n$ is approximately $\sqrt{n}$ times narrower than $U$ itself. For a single emitter ($N = 1$), just raising the signal $F(\mathbf{r},t)$ to certain power $n > 1$ would be sufficient for achieving the effect. For $N > 1$ emitters, exponentiation of the signal leads to additional cross-terms, hindering resolution of the emitters. For example, for $n = 2$ the squared signal equals
\begin{multline}
    \label{eq:signal_squared}
    F^2(\mathbf{r}, t) = \sum_k U^2(\mathbf{r} - \mathbf{r}_k) \varepsilon_k^2 s_k^2(t) \\ + \sum_{k, k': k' \ne k} U(\mathbf{r} - \mathbf{r}_k) U(\mathbf{r} - \mathbf{r}_{k'}) \varepsilon_k \varepsilon_{k'} s_k(t) s_{k'}(t).
\end{multline}
To ensure resolution enhancement, one needs to construct a polynomial function of the fluorescence signal with excluded cross-terms \cite{dertinger2009fast}. For example, variance of the signal considered as a time series of frames both is quadratic relatively to PSF and has zero cross-terms for independent blinking of the emitters:
\begin{equation}
    \label{eq:SOFI_variance}
    \operatorname{Var}_t [F(\mathbf{r}, t)] = \sum_k U^2(\mathbf{r} - \mathbf{r}_k) \varepsilon_k^2 \operatorname{Var}_t[s_k(t)].
\end{equation}

The approach can be extended to higher-order polynomials by using cumulants instead of variance \cite{dertinger2009fast}:
\begin{equation}
    \label{eq:SOFI_cumulant}
    C_t^{(n)} [F(\mathbf{r}, t)] = \sum_k U^n(\mathbf{r} - \mathbf{r}_k) \varepsilon_k^n C_t^{(n)}[s_k(t)].
\end{equation}
The expression stems from linearity of cumulants relatively to a sum of independent random variables: $C^{(n)}[X_1 + X_2] = C^{(n)}[X_1] + C^{(n)}[X_2]$ for independent $X_1$ and $X_2$.

Several lower-order cumulants for a random variable (process) $f(t)$ are defined as
\begin{equation}
    \label{eq:c1_def}
    C_t^{(1)}[f(t)] = \operatorname{E}_t[f(t)],
\end{equation}
\begin{equation}
    \label{eq:c2_def}
    C_t^{(2)}[f(t)] = \operatorname{E}_t\bigl[\bigl(f(t) - \operatorname{E}_t[f(t)]\bigr)^2 \bigr],
\end{equation}
\begin{equation}
    \label{eq:c3_def}
    C_t^{(3)}[f(t)] = \operatorname{E}_t\bigl[\bigl(f(t) - \operatorname{E}_t[f(t)]\bigr)^3 \bigr],
\end{equation}
\begin{equation}
    \label{eq:c4_def}
    C_t^{(4)}[f(t)] = \operatorname{E}_t\bigl[\bigl(f(t) - \operatorname{E}_t[f(t)]\bigr)^4 \bigr] - 3 \bigl( C_t^{(2)}[f(t)] \bigr)^2,
\end{equation}
and correspond to variance, asymmetry, and excess kurtosis respectively. Here, $\operatorname{E}_t[f(t)]$ denotes time averaging of the random values $f(t)$ (defined at each frame $t$) over an infinite number of frames $N_\text{f}$. The random process of blinking is assumed to be stationary (with negligible effect of bleaching) so that ergodicity can be applied to the definition of cumulants. For finite acquisition time, expectation values are estimated as finite-sample averages:
\begin{equation}
    \label{eq:finite_sample}
    \operatorname{E}_t[f(t)] \approx \frac{1}{N_\text{f}} \sum_{t = 1}^{N_\text{f}}f(t).
\end{equation}

Further, one can define also cross-cumulants and introduce non-zero time lags \cite{dertinger2010achieving}:
\begin{multline}
    \label{eq:SOFI_cross-cumulant}
    C_t^{(n)} [F(\mathbf{r}^{(1)}, t+\tau_1), \ldots, F(\mathbf{r}^{(n)}, t+\tau_n)] \\ \propto \sum_k U^n(\bar{\mathbf{r}} - \mathbf{r}_k) \varepsilon_k^n C_t^{(n)}[s_k(t+\tau_1), \ldots, s_k(t+\tau_n)],
\end{multline}
where $\bar{\mathbf{r}} = \sum_{i=1}^n \mathbf{r}^{(i)} / n$. 

To increase spatial resolution, SOFI deals with stochastic fluctuations of the \textit{total} fluorescence signal, integrated over each frame duration and certain spectral range, rather than on spectral or temporal intensity distribution \textit{inside} a frame. The technique itself does not impose any specific constraints on time or spectral windows, used during signal acquisition \textit{within} frames. That property is crucial for the proposed sensing approach and was efficiently used in the color SOFI \cite{gallina2013resolving,glogger2021multi}. It is worth noting, that there also exists an alternative approach of examining the emitters' surrounding by analysis of the parameters of fluctuation \textit{already used} for SOFI (for example, on-time ratio or blinking frequency) \cite{geissbuehler2012mapping,merdasa2017supertrap,mo2017genetically} rather than additional ``degrees of freedom''.

\color{black}

\subsection{Limitations and extensions of SOFI}
\label{subsec:SOFI limitations}

Derivation of the cornerstone Eq.~(\ref{eq:SOFI_cumulant}) in the classical version of SOFI \cite{dertinger2009fast,dertinger2010achieving,dertinger2013advances} assumes that the signal fluctuations are dominated by the emitters' blinking, while the quantum shot noise effects can be neglected. The assumption is fully legitimate in the high-brightness limit, but causes artifacts in zero-time-lag autocumulant images for finite numbers of detected photons \cite{dertinger2009fast,pawlowska2021embracing}. In a few-photon regime, the shot noise prevails and completely hinders super-resolving capabilities of SOFI \cite{picariello2025imaging}. 

To resolve the issue and extend SOFI applicability to low-light imaging, F.~Picariello and coauthors \cite{picariello2025imaging} proposed to replace SOFI cumulants (see Eqs.~(\ref{eq:c2_def})--(\ref{eq:c4_def})) by their linear combinations ensuring cancellation of the shot noise contributions. The approach has been termed quantum super-resolution imaging by photon statistics (QSIPS) \cite{picariello2025imaging}. It restores validity of Eq.~(\ref{eq:SOFI_cumulant}) for any intensity level.

Later, it has been noticed \cite{peshko2025up} that the introduced cumulant combinations represent factorial cumulants \cite{kendall1977advanced} and can be easily obtained from normal SOFI cumulants by replacing raw moments $\operatorname{E}_t[f^n(t)]$ by the corresponding factorial moments
\begin{equation}
    \label{eq:factorial_moments}
    \operatorname{E}_t[f^n(t)] \mapsto \operatorname{E}_t[f(t)\{f(t)-1\}\cdots \{f(t) - n + 1\}],
\end{equation}
where it is explicitly assumed that the signal $f(t)$ describes the number of photons detected in the frame $t$. From quantum optics perspective, the transition from SOFI to QSIPS can be understood as imposing normal ordering to the field operator moments, which is quite natural for description of light detection in a few-photon regime \cite{peshko2025up}.

Another important limitation of SOFI applicability is related to the scaling of signal-to-noise ratio (SNR) with the density of emitters distribution over the sample. The useful SOFI (and QSIPS) signal is given by Eq.~(\ref{eq:SOFI_cumulant}), which implies its scaling as $C_t^{(n)}[F(\mathbf{r},t)] \propto M$ with the number of emitters $M$ contributing to the detection point $\mathbf{r}$. The statistical noise is accumulated over all terms of the expression actually used for construction of the cumulant (see Eqs.~(\ref{eq:c2_def})--(\ref{eq:c4_def})). It can be estimated as $\Delta C_t^{(n)}[F(\mathbf{r},t)] \propto \sqrt{\operatorname{E}_t[F^n(t)]} \propto M^{n/2}$ by assuming Poisson statistics of the intensity noise and taking into account that contributions of separate emitters are added up: $F(t) \propto M$. The resulting SNR scaling takes the form
\begin{equation}
    \label{eq:SNR_scaling}
    \text{SNR} \equiv \frac{C_t^{(n)}[F(\mathbf{r},t)]}{\Delta C_t^{(n)}[F(\mathbf{r},t)]} \propto M^{1-(n/2)}.
\end{equation}

The number of emitters $M$ contributing to a specific detection point can be estimated as $M \sim \rho_\text{2D} d_\text{R}^2$ for 2-dimensional arrangement of emitters with the surface density $\rho_\text{2D}$ (the number of emitters per unit area), where the classical Rayleigh resolution limit $d_\text{R}$ defines the characteristic width of the microscope PSF. For 1-dimensional (filament-like) arrangement of emitters with the linear density $\rho_\text{1D}$ (the number of emitters per unit length), the number $M$ scales as $M \sim \rho_\text{1D} d_\text{R}$.

For the traditional intensity measurement ($n = 1$), higher surface or linear density $\rho$ of emitters leads to a brighter image and better SNR: $\text{SNR} \propto \sqrt{\rho}$. The 2nd order cumulant does not demonstrate pronounced change of the SNR with the emitters density. Further increase of the cumulant order leads to adverse scaling of the SNR with the emitters density: for example, the 4th order cumulant behaves as $\text{SNR} \propto \rho^{-1}$ and accumulates significant contribution of shot noise for a dense configuration of emitters. 

It is also instructive to express the number of emitters $M$ in the PSF spot in terms of the typical distance $d$ between adjacent emitters: $M \sim (d_\text{R} / d)^2$ for 2-dimensional arrangement and $M \sim d_\text{R} / d$ for a filament-like structure. The same performance of SOFI can be achieved for much closer emitters in a 1-dimensional structure relatively to a 2-dimension one. For example, $M = 25$ corresponds to $d \sim d_\text{R} / 25$ and $d \sim d_\text{R} / 5$ for 1- and 2-dimensional arrangements respectively.

\color{black}

\subsection{Sensing with fluorescent nanoparticles}

\begin{figure}[tp]
    \centering
    \includegraphics[width=\linewidth]{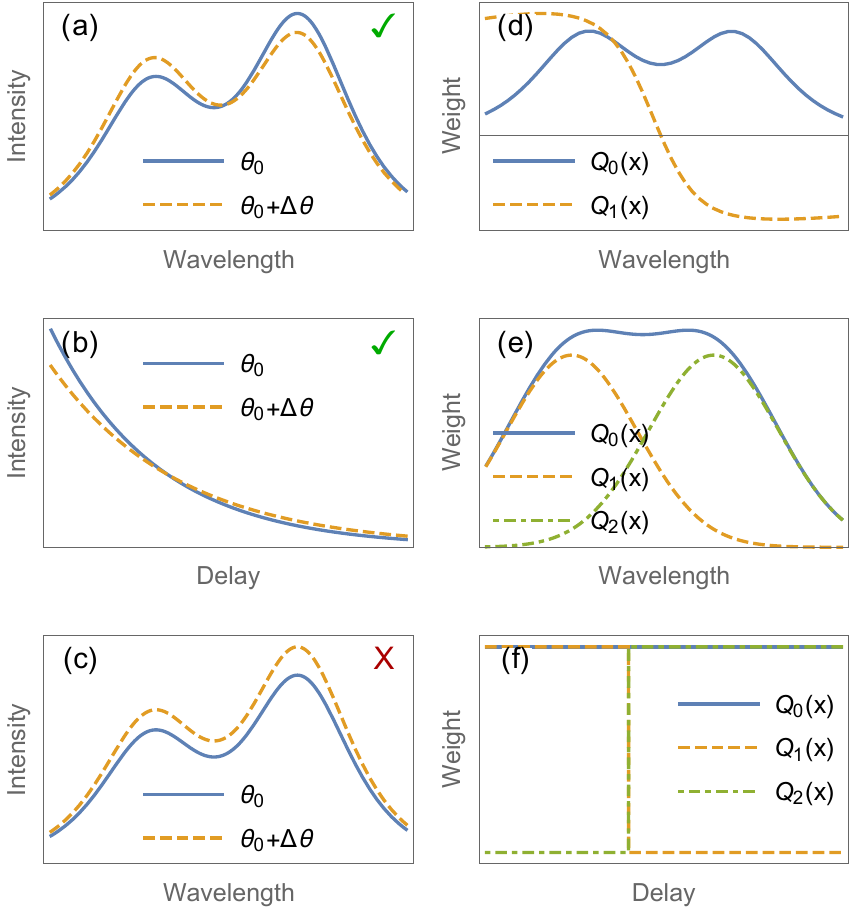}
    \caption{Sensing an environment parameter $\theta$ by its influence on fluorescence signal. The value of the parameter can be inferred from changes of the spectrum (a), fluorescence lifetime (b), or total intensity (c). The latter case is likely to yield worse sensing accuracy because of blinking-induced noise of absolute intensity \cite{geissbuehler2012mapping}. The full spectral profile (a) can be compressed into two channels: with the optimal weights described in Appendix~\ref{app:optimal_estimator} (d) and with practical spectral windows having Gaussian profiles of sensitivity $Q_1(x)$ and $Q_2(x)$ (e). The channel $Y_0$ is defined as $Y_1 + Y_2$ and corresponds to the weight $Q_0(x) = Q_1(x) + Q_2(x)$. The full temporal profile (b) can be compressed into two channels by using a simple threshold filter (f).}
    \label{fig:sensing}
\end{figure}

Sensing capabilities of fluorescent nanoparticles can be provided by dependence of their fluorescence lifetime, spectrum, or total intensity on the local parameters of the environment: temperature, pH, pressure, etc. \cite{norman2021novel,wu2022recent,bushuev2016synthesis,maletinsky2012robust,arcizet2011single,babinec2010diamond,lesik2013maskless,golubewa2022all,Galland2011lifetime,qu2013ratiometric,wang2014carbon} (Fig.~\ref{fig:sensing}). For example, carbon nanodots possess spectral dependence of fluorescence on pH and temperature \cite{qu2013ratiometric} and can exhibit blinking required for SOFI \cite{Chizhik2016CND}. Surface-induced charge-state switching of color centers in nanodiamonds \cite{petrakova2015charge,rondin2010surface,aslam2013photo,siyushev2013optically,han2017surface,alkahtani2020charge} enables combination of their vast sensing capabilities \cite{norman2021novel,wu2022recent,bushuev2016synthesis,maletinsky2012robust,arcizet2011single,babinec2010diamond,lesik2013maskless} with SOFI. 

In other words, variation of the parameter of interest can change the \textit{total} intensity of the fluorescent signal and (or) its \textit{distribution} over certain parameter (variable) $x$ (wavelength, emission delay, etc.) within the frame. Variations of molecular surrounding \cite{bruni2017two,schaferling2012art,demchenko2009monitoring} and blinking of the emitters (required for SOFI applicability) \cite{geissbuehler2012mapping} introduce significant noise to the total (absolute) intensity of fluorescence. For that reason, further we focus on the latter case when the sensed environmental parameter influences the \textit{distribution} of the fluorescent signal. Let $f(\mathbf{r}, t, x)$ be the density of the signal over the variable $x$, connected with the total fluorescent signal $F(\mathbf{r},t)$ for the frame $t$ by the relation
\begin{equation}
    F(\mathbf{r},t) = \int f(\mathbf{r}, t, x) dx.
\end{equation}
For spectral response of the fluorophores to local changes of environment, $x$ may correspond to the emitted radiation wavelength and $f(\mathbf{r}, t, x)$ be the spectral density of the signal. For sensing, based on fluorescence lifetime like in the case of SiV centers in diamond needles \cite{golubewa2022all}, $x$ is the delay between the excitation pulse and the detection of photons.

Eq.~(\ref{eq:basic_signal}) can be rewritten in the following form for the density of the signal:
\begin{equation}
    \label{eq:density_signal}
    f(\mathbf{r}, t, x) = \sum_k U(\mathbf{r} - \mathbf{r}_k) \varepsilon_k(x, \theta_k) s_k(t),
\end{equation}
where the dependence on the $k$-th emitter intensity $\varepsilon_k(x,\theta)$ on the variable $x$ (wavelength or delay of the fluorescence signal) is affected by the local value of the environment parameter $\theta = \theta_k$ in the vicinity of the $k$-th emitter. To infer the local values $\theta(\mathbf{r})$ of the parameter in question, one needs to specify certain model for the dependence of $\varepsilon_k(x, \theta)$ on $\theta$. Typically, the emitters are assumed to be described by the same model $\varepsilon(x, \theta_k)$ up to a constant emitter-dependent brightness factor $\varepsilon_k$: i.e., $\varepsilon_k(x, \theta_k) = \varepsilon_k \varepsilon(x, \theta_k)$.

When trying to combine SOFI {\color{black}(or QSIPS)} with sensing, it is tempting to apply cumulants construction to each value of $x$ (spectral or temporal channel) independently. However, as we discuss in the next section, splitting the fluorescent signal into a large number of channels leads to low photon numbers, significant shot noise, and inefficiency of high-order cumulants usage in SOFI. On the other hand, reconstruction of the whole distribution $\varepsilon(x,\theta(\mathbf{r}))$ is not needed to infer the spatial map of a single parameter $\theta(\mathbf{r})$. As we show here, the whole intensity profile can be ``compressed'' to just two scalar values with their ratio being sufficient for estimation of the parameter $\theta$. Accumulation of the intensity into two such channels effectively reduces the shot noise and restores applicability of SOFI for spatial super-resolution.

For now, let us consider sensing by a single probe nanoparticle localized at the position $\mathbf{r}_0$ and subjected to the environment with the parameter value $\theta$:
\begin{equation}
    \label{eq:density_signal_single}
    f(\mathbf{r}, t, x) = U(\mathbf{r} - \mathbf{r}_0) \varepsilon_0(x, \theta) s_0(t).
\end{equation}

To eliminate additional inaccuracies, caused by blinking of the emitter, finite data acquisition time, or unknown brightness $\varepsilon_0$, one can work with normalized signal by dividing its density over the total value:
\begin{equation}
    \label{eq:normalized signal}
    \tilde f(\mathbf{r},t,x) \equiv \frac{f(\mathbf{r},t,x)}{\int f(\mathbf{r},t,x) dx} = \frac{\varepsilon(x, \theta)}{\int \varepsilon(x, \theta) dx}.
\end{equation}
The idea of analyzing ratios instead of absolute values of the signal for reduction of fluctuations effects is widely used in ratiometric imaging \cite{bruni2017two,klymchenko2022fluorescent} {\color{black}and in FRET-SOFI \cite{valenta2025super}}. One can notice that the normalized signal depends on the normalized density of the emitter's signal and is determined by the parameter of interest $\theta$, but does not include irrelevant geometrical information or the emitter brightness.

Instead of considering the whole normalized profile $\tilde f(\mathbf{r}_0,t,x)$ for a specific frame $t$ and the emitter position $\mathbf{r}_0$, one can construct a quantity
\begin{equation}
    \label{eq:Z definition}
    Z = \frac{Y_1}{Y_0},\quad Y_i = \int f(\mathbf{r}_0, t, x) Q_i(x)  dx
\end{equation}
for certain weight functions $Q_0(x)$ and $Q_1(x)$ and consider it as a ``compressed'' version of the measured signal (Fig.~\ref{fig:sensing}(d-f)). As shown in Appendix~\ref{app:optimal_estimator}, there exists the (locally) optimal choice of the weights (Fig.~\ref{fig:sensing}(d)), for which the quantity $Z$ contains exactly the same amount of information about the sensed parameter $\theta$ as the whole intensity profile $\tilde f(\mathbf{r}_0,t,x)$ (i.e., represents \textit{sufficient statistics}).

From practical point of view, Eq.~(\ref{eq:Z definition}) defines two detection channels $Y_0$ and $Y_1$ and proposes to use their ratio $Z$ for inference of $\theta$. In a real experiment, implementation of the optimal detection channels (with the optimal weight functions $Q_0(x)$ and $Q_1(x)$) can be complicated or impossible. Alternatively, one can just use the signals from two practically available channels $Y_1$ and $Y_2$ with their weight functions $Q_1(x)$ and $Q_2(x)$ (for example, corresponding to the signal accumulated over two spectral or temporal windows --- Fig.~\ref{fig:sensing}(e,f)), define $Y_0 = Y_1 + Y_2$ (and $Q_0(x) = Q_1(x) + Q_2(x)$), and construct $Z$ according to Eq.~(\ref{eq:Z definition}). Strictly speaking, such measurement-induced choice of the weights is, generally, non-optimal and may lead to loss of information relatively to the full profile analysis.

Substituting Eq.~(\ref{eq:normalized signal}) into Eq.~(\ref{eq:Z definition}), one can find the relation between the parameter $\theta$ and the compressed signal $Z$:
\begin{equation}
    \label{eq:Z vs theta}
    Z = \frac{ \int \varepsilon(x, \theta) Q_1(x) dx }{\int \varepsilon(x, \theta) Q_0(x) dx} \equiv \zeta(\theta).
\end{equation}
If the \textit{a priori} constraints $\theta_\text{min} \le \theta \le \theta_\text{max}$ on the sensed $\theta$ ensure monotonicity of the dependence $\zeta(\theta)$, the function $\zeta(\theta)$ can be inverted and the estimator (mapping of the signal to the parameter of interest) can be constructed as
\begin{equation}
    \label{eq:estimator}
    \bar \theta = \zeta^{-1}(Z).
\end{equation}

For a particular case of linear dependence of the emitter's signal density on the parameter of environment in the region of interest $[\theta_\text{min}, \theta_\text{max}]$
\begin{equation}
    \label{eq:density_profile}
    \varepsilon(x, \theta) = S_0(x) + S_1(x) \theta,
\end{equation}
the estimator can be constructed analytically:
\begin{equation}
    \label{eq:linear_estimator_practical}
    Z = \frac{A_{10} + A_{11} \theta}{A_{00} + A_{01} \theta} \quad \Rightarrow \quad
    \bar \theta = \frac{A_{00} Z - A_{10}}{A_{11} - A_{01} Z},
\end{equation}
where
\begin{equation}
    \label{eq:constant_A}
    A_{ij} = \int Q_i(x) S_j(x)  dx, \quad i,j = 0, 1.
\end{equation}

It is worth mentioning that for the optimal weights $Q_0(x)$ and $Q_1(x)$, defined in Appendix~\ref{app:optimal_estimator}, we have $A_{ij} = \delta_{ij}$ and $\bar \theta = Z$.

\subsection{Combining sensing and SOFI}

The proposed idea of combining spatial resolution with sensing consists in applying SOFI {\color{black}(QSIPS)} separately to several channels relatively to the variable $x$, which defines signal distribution (spectral or temporal) within each frame. Naively, one can just replace the integrated signal $F(\mathbf{r},t)$ by its density $f(\mathbf{r}, t, x)$ in Eq.~(\ref{eq:SOFI_cumulant}) and get spectrally or temporarily resolved cumulant image:
\begin{equation}
    \label{eq:SOFI_cumulant_x}
    C_t^{(n)} [f(\mathbf{r}, t, x)] = \sum_k U^n(\mathbf{r} - \mathbf{r}_k) \varepsilon_k^n \varepsilon^n(x, \theta_k) C_t^{(n)}[s_k(t)].
\end{equation}
Practically, such approach would mean that one takes the binned signal density $f(\mathbf{r}, t, x_i)$, specified for a discrete set of values $x_i$, and performs analysis of frame-wise temporal statistics for each bin $x_i$ by considering the time series $\{f(\mathbf{r}, t, x_i\}_t$. However, such time series is {\color{black}likely to have a small number of photons, leading to inapplicability of classical SOFI and to large statistical inaccuracy for higher-order cumulants in QSIPS.}

Such issue can be resolved by the above-discussed ``compression'' of a full profile over $x$ into several scalar quantities. Following Eq.~(\ref{eq:Z definition}), one can define the signals, integrated over two spectral or temporal windows (indexed as $j = 0, 1$) within each frame, as follows:
\begin{multline}
    F_j(\mathbf{r}, t) = \int f(\mathbf{r}, t, x) Q_j(x) dx  \\ = \sum_k U(\mathbf{r} - \mathbf{r}_k) \varepsilon_k  Y_j(\theta_k) s_k(t), \quad j = 0, 1,
\end{multline}
where 
\begin{equation}
    Y_j(\theta_k) = \int \varepsilon(x, \theta_k) Q_j(x) dx
\end{equation}
is the $j$-th channel signal of the $k$-th emitter subjected to the environmental parameter value $\theta = \theta_k \equiv \theta(\mathbf{r}_k)$. 

In contrast to the binned density $f(\textbf{r}, t, x_i)$, the integrated signals $F_j(\mathbf{r}, t)$ are intense enough to be exploited for SOFI {\color{black}or to yield reliable enough signal in QSIPS}. The cumulants can be constructed for each window separately:
\begin{multline}
    \label{eq:cumulant_for_window}
    C_j^{(n)}(\mathbf{r}) = C_t^{(n)}[F_j(\mathbf{r}, t)] \\ = \sum_k U^n(\mathbf{r} - \mathbf{r}_k) \varepsilon_k^n  Y_j^n(\theta_k) C_t^{(n)}[s_k(t)].
\end{multline}
In the way, inherent to SOFI, the blurring is reduced by raising the PSF $U$ to the power $n$. The expressions for the quantities $C_0^{(n)}(\mathbf{r})$ and $C_1^{(n)}(\mathbf{r})$ differ only by the factors $Y_0^n(\theta_k)$ and $Y_1^n(\theta_k)$ in the sums, while the leading contribution to the sum is provided by the emitters with $\mathbf{r}_k \approx \mathbf{r}$. Therefore, their ratio can be considered as an approximation of the local value of the compressed signal $Z(\mathbf{r})$:
\begin{equation}
    \label{eq:Z_for_cumulants}
    Z(\mathbf{r}) \equiv \zeta(\theta(\mathbf{r})) \approx Z_n(\mathbf{r}) \equiv \left[\frac{C_1^{(n)}(\mathbf{r})}{C_0^{(n)}(\mathbf{r})}\right]^{1/n}.
\end{equation}
For even $n$, the information about the sign of $Z(\mathbf{r})$ can be lost, since $Y_j^n(\theta_k) \ge 0$ for even $n$ regardless of $\operatorname{sign}Y_j(\theta_k)$. However, that limitation is typically irrelevant in practice, because non-negative weights $Q_j(x) \ge 0$ for integration over spectral or temporal windows ensure $Y_j \ge 0$.

The resulting estimator for the $n$-th order SOFI combined with sensing is 
\begin{equation}
    \label{eq:estimator_SOFI}
    \bar \theta^{(n)}(\mathbf{r}) = \zeta^{-1} \left(Z_n(\mathbf{r})\right).
\end{equation}

\color{black}

It is important that the construction of estimator $\bar \theta^{(n)}(\mathbf{r})$ does not rely on any prior knowledge about the emitters' parameters: their molecular brightness $\varepsilon_k$ or blinking kinetics, which determines the cumulant values $C_t^{(n)}[s_k(t)]$. In Eq.~(\ref{eq:Z_for_cumulants}), those quantities effectively cancel out, their variation between emitters leading to unequal (weighted) contributions to the estimated parameter values (see also Eq.~(\ref{eq:estimator linear approx.}) below).

\color{black}

The estimator possesses the following properties justifying its application (see Appendix~\ref{app:cumulant-based_estimator}):
\begin{enumerate}
    \item For a uniform (constant) parameter distribution $\theta(\mathbf{r}) = \theta_0$, the estimator reproduces the correct parameter value for all orders $n$:
    \begin{equation}
        \label{eq:estimator uniform}
        \bar \theta^{(n)}(\mathbf{r}) = \theta_0 \text{ for all $\mathbf{r}$ and $n$}. 
    \end{equation}
    \item If all the emitters are subjected to the environment parameter $\theta$ limited to certain range: $\theta_k \in [\theta_\text{min}, \theta_\text{max}]$ for all $k$, the estimator will also belong to that range:
    \begin{equation}
        \bar \theta^{(n)}(\mathbf{r}) \in [\theta_\text{min}, \theta_\text{max}]\text{ for all $\mathbf{r}$ and $n$}.
    \end{equation}
    \item For small variations of the parameter $\theta$ (i.e., in the linear approximation), the estimator will reproduce a weighted combination of the parameter values $\theta_k$ for the emitters in the vicinity of the point of interest:
    \begin{equation}
        \label{eq:estimator linear approx.}
        \bar \theta^{(n)}(\mathbf{r}) \approx \frac{\sum_k U^n(\mathbf{r} - \mathbf{r}_k)C_t^{(n)}[s_k(t)]  \varepsilon_k^n \theta_k}{\sum_k U^n(\mathbf{r} - \mathbf{r}_k) C_t^{(n)}[s_k(t)] \varepsilon_k^n}.
    \end{equation}
    For larger $n$, the estimate becomes more localized due to raising the PSF $U$ to the power $n$.
\end{enumerate}

\subsection{Reconstruction of parameter distribution for noisy signal}
\label{subsec:noisy}

While for ideal noiseless data the constructed estimator (\ref{eq:estimator_SOFI}) would provide reasonable local values for the investigated parameter distribution $\theta(\mathbf{r})$, its application to noisy data (especially for sparsely located emitters) deserves additional attention. The structure of expression~(\ref{eq:cumulant_for_window}) implies that the cumulant values $C_j^{(n)}(\mathbf{r})$ estimated from finite-sample data are more reliable near emitters ($\mathbf{r} \approx \mathbf{r}_k$ for certain $k$), but may contain significant relative shot noise contribution in the region between the emitters where the signals are weak. The effect becomes more pronounced for higher-order cumulants, since the effective PSF $U^n(\mathbf{r} - \mathbf{r}_k)$ is narrower for larger $n$, while the contribution of shot noise is generally growing for larger cumulant orders.

The inaccuracy $\Delta \bar \theta^{(n)} (\mathbf{r})$ of the reconstructed parameter value $\bar \theta^{(n)} (\mathbf{r})$ is determined by the relative contribution of the shot noise to the quantities $C_0^{(n)}(\mathbf{r})$ and $C_1^{(n)}(\mathbf{r})$, which is, roughly, inversely proportional to the square root of the detected signal:
\begin{equation}
    \Delta \bar \theta^{(n)} (\mathbf{r}) \propto \frac{1 }{\sqrt{\left|C_0^{(n)}(\mathbf{r}) + C_1^{(n)}(\mathbf{r})\right|}}.
\end{equation}
The likelihood of getting the estimated distribution $\bar \theta^{(n)} (\mathbf{r})$ for the true distribution $\theta(\mathbf{r})$ can be represented as
\begin{equation}
    L\bigl[\bar \theta^{(n)}(\mathbf{r}) \big| \theta(\mathbf{r})\bigr] \propto \exp \left( - \int d^2 \mathbf{r} \frac{\left[\bar \theta^{(n)}(\mathbf{r}) - \theta(\mathbf{r})\right]^2}{2 \left[\Delta \bar \theta^{(n)}(\mathbf{r})\right]^2} \right).
\end{equation}

In the regions with low signal, the inaccuracy $\Delta \bar \theta (\mathbf{r})$ can be large, making the estimate $\bar \theta (\mathbf{r})$ unreliable. The estimate can be refined in the spirit of Tikhonov's regularization \cite{tikhonov1977solutions}. Very sharp features of the reconstructed distribution of the environment parameter $\theta$ from a blurred image with finite density of fluorescent emitters are likely to be artifacts rather than to provide useful information about the sample. The statement can be captured as a reduced prior probability of the distributions with large values of second derivatives:
\begin{equation}
    \label{eq:regularizing_p0}
    p_0[\theta(\mathbf{r})] \propto \exp \left( \frac{1}{2 D^2}\int d^2 \mathbf{r} \left| \nabla^2  \theta(\mathbf{r}) \right|^2 \right),
\end{equation}
where the parameter $D$ represents the scale for the derivatives and $\nabla^2 = \partial^2 / \partial x^2 + \partial^2 / \partial y^2$ is 2-dimensional Laplacian.

According to Bayes' formula \cite{gelman2013,koch2007}, the posterior probability of the true distribution $\theta(\mathbf{r})$ conditioned by the estimated distribution $\bar \theta^{(n)}(\mathbf{r})$ equals
\begin{equation}
    \label{eq:p_regularization}
    p\bigl[\theta(\mathbf{r}) \big| \bar \theta^{(n)}(\mathbf{r})\bigr] \propto L\bigl[\bar \theta^{(n)}(\mathbf{r}) \big| \theta(\mathbf{r})\bigr] p_0[\theta(\mathbf{r})].
\end{equation}
The distribution $\theta(\mathbf{r})$ can be reconstructed according to maximum \textit{a posteriori} method:
\begin{equation}
    \label{eq:theta_corrected}
   \theta(\mathbf{r}) = \argmax\limits_{\theta(\mathbf{r})} p\bigl[\theta(\mathbf{r}) \big| \bar \theta^{(n)}(\mathbf{r})\bigr].
\end{equation}

Practically, the parameter values are estimated at a discrete grid of points $\mathbf{r}_{ij}$. In that case, the derivatives in Eq.~(\ref{eq:p_regularization}) can be replaced by finite differences, and the values $\theta(\mathbf{r}_{ij})$ can be found by unconstrained quadratic minimization (Appendix~\ref{app:regularized_expressions}).

\subsection{Sensing method summary}

To realize the proposed sensing approach with spatial super-resolution capabilities, the following procedures should be performed.

1. Model construction. The model for parameter-dependent spectral or temporal profiles $\varepsilon(x, \theta)$ of the emitters' fluorescence should be constructed (either theoretically or from empirical calibration procedure).

2. Sensing calibration. The optimal weights $Q_0(x)$ and $Q_1(x)$ of the two detection channels are defined in Appendix~\ref{app:optimal_estimator}. If the experimental capabilities do not allow to perform such optimal measurement, the actual weights $Q_0(x)$ and $Q_1(x)$ are defined by the available detection setup (the profiles of the channels' sensitivity). Based on the functions $\varepsilon(x, \theta)$, $Q_0(x)$, and $Q_1(x)$, the dependence $\zeta(\theta)$ of the compressed signal on the sensed parameter should be constructed according to Eq.~(\ref{eq:Z vs theta}). Then, the estimator for the $n$-th order of SOFI {\color{black}or QSIPS} is described by Eq.~(\ref{eq:estimator_SOFI}).

3. Data acquisition. The fluorescent response of the sample is collected in terms of 2-channel framed images $F_k(\mathbf{r}_{ij}, t)$, where $k = 0$, 1 is the channel index, $t$ is the frame index, and $\{ \mathbf{r}_{ij} \}$ is the set of detection positions (pixel positions for an array detector or the scanning positions).

4. Data processing. For each channel $k$ and each detection position $\mathbf{r}_{ij}$ separately, the cumulant value $ C_k^{(n)}(\mathbf{r}_{ij}) = C_t^{(n)}[F_k(\mathbf{r}_{ij}, t)]$ is calculated. For $n = 2$, 3, 4 the expressions are provided by Eqs.~(\ref{eq:c2_def})--(\ref{eq:c4_def}) with the expectation values replaced by finite-sample averages. Alternatively, one can use cross-cumulants for improvement of SNR as described in Appendix~\ref{app:cross-cumulants}. {\color{black}For low-brightness regime, SOFI cumulants should be replaced by QSIPS cumulants (or, equivalently, the raw moments should be replaced by factorial moments according to Eq.~(\ref{eq:factorial_moments})). The SNR improvement by using cross-cumulants is applicable to QSIPS as well \cite{peshko2025up}.} The local estimates of the parameter $\bar \theta^{(n)}(\mathbf{r})$ are calculated according to Eq.~(\ref{eq:estimator_SOFI}).

5. Data interpolation. To account for the difference of the estimate reliability at different points, the corrected parameter distribution is constructed according to Eq.~(\ref{eq:theta_corrected}) with the method described in Appendix~\ref{app:regularized_expressions}.

\section{Numerical modeling}
\label{sec:modeling}

\subsection{Modeled setup}

To illustrate the proposed combined sensing approach, we apply it to simulated datasets. The modeled objects include regular 2-dimensional grids of fluorescent probes (Subsections~\ref{subsec:high-brightness} and \ref{subsec:low-brightness}) or a set of 1-dimensional filament-like structures (Subsection~\ref{subsec:1D structure}) with the distance $d$ between adjacent emitters smaller that the Rayleigh resolution limit $d_\text{R}$ of the microscope for classical incoherent-light imaging. Each probe is assigned certain ``true'' value $\theta_k \in [0, 1]$ of the sensed parameter $\theta$ according to the modeled spatial distribution.

The blinking of the modeled emitters is described by the telegraph process with the waiting time for switching between the bright (``on'') and dark (``off'') states distributed according to the probability densities
\begin{equation}
    p_\text{on,off}(\Delta t) = t_\text{on,off} \exp\left(- \Delta t / t_\text{on,off}\right),
\end{equation}
where $t_\text{on}$ and $t_\text{off}$ are the mean durations of continuous staying in the bright and dark state respectively. The natural time unit for the modeling is the SOFI frame duration. In those units, the ``on'' and ``off'' times used for the modeling are $t_\text{on} = 2$~frames and $t_\text{off} = 3$~frames. We assume that the bright state emission is described by Poisson distribution of the photon numbers.

According to the considered model, the fluorescence signal is measured by a $40 \times 40$ pixels array detector. {\color{black} For demonstration purposes, we consider high and low-brightness regimes with the average number of photons detected from each emitter (in total over the whole detection plane) equal to $\langle N \rangle = (1 - 4)\times 10^4$ and $\langle N \rangle = 4$ photons per frame respectively. The high-brightness regime is typical for fluorescent quantum dots with their emission rates ranging from $10^3$ to $10^5$ photons per emitter per frame \cite{dertinger2009fast,sroda2020sofism,kurdzialek2021super,ye2011blinking}. Consideration of the low-brightness regime demonstrates applicability of the developed approach to low-light imaging \cite{picariello2025imaging} and extends the range of acceptable emission rates below the values $10^2 - 10^3$ photons per emitter per frame characteristic for organic dyes \cite{grussmayer2020spectral}, florescent proteins \cite{vandenberg2019extended}, and blinking color centers in nanodiamond \cite{bradac2010observation,aslam2013photo}. The total observation time is assumed to include $1.5\times 10^{4}$~frames for the high-brightness regime (Subsections~\ref{subsec:high-brightness} and \ref{subsec:1D structure}) and $1.5\times 10^{6}$~frames for the low-brightness regime (Subsection~\ref{subsec:low-brightness}).} 

The signal is split into two spectral channels (Fig.~\ref{fig:sensing}(e)) in the ratio 
\begin{equation}
    Y_1 : Y_2 = (0.75 + 0.05 \theta) : (1.25 - 0.05 \theta).
\end{equation}
The signal $Y_0$, used for normalization in denominator of Eq.~(\ref{eq:Z definition}), is defined as the sum of both channels: $Y_0 = Y_1 + Y_2$. Here, we assume that the dependence of the emitters' response to the variations of the sensed parameter $\theta$ is linear and follows Eq.~(\ref{eq:density_profile}). According to the ratios of the channel signals, defined above, the coefficients in Eq.~(\ref{eq:linear_estimator_practical}) are related as $A_{01} = 0$, $A_{10} = 0.375 A_{00}$, and $A_{11} = 0.025 A_{00}$. Therefore, the estimator (\ref{eq:linear_estimator_practical}) takes the form
\begin{equation}
    \bar \theta = \zeta^{-1}(Z) = 40 Z - 15.
\end{equation}

The source code for the discussed numerical modeling is available in Supplemental Material at [URL will be inserted by publisher]. 

\subsection{High-brightness regime}
\label{subsec:high-brightness}

\begin{figure}[tp]
    \centering
    \includegraphics[width=\linewidth]{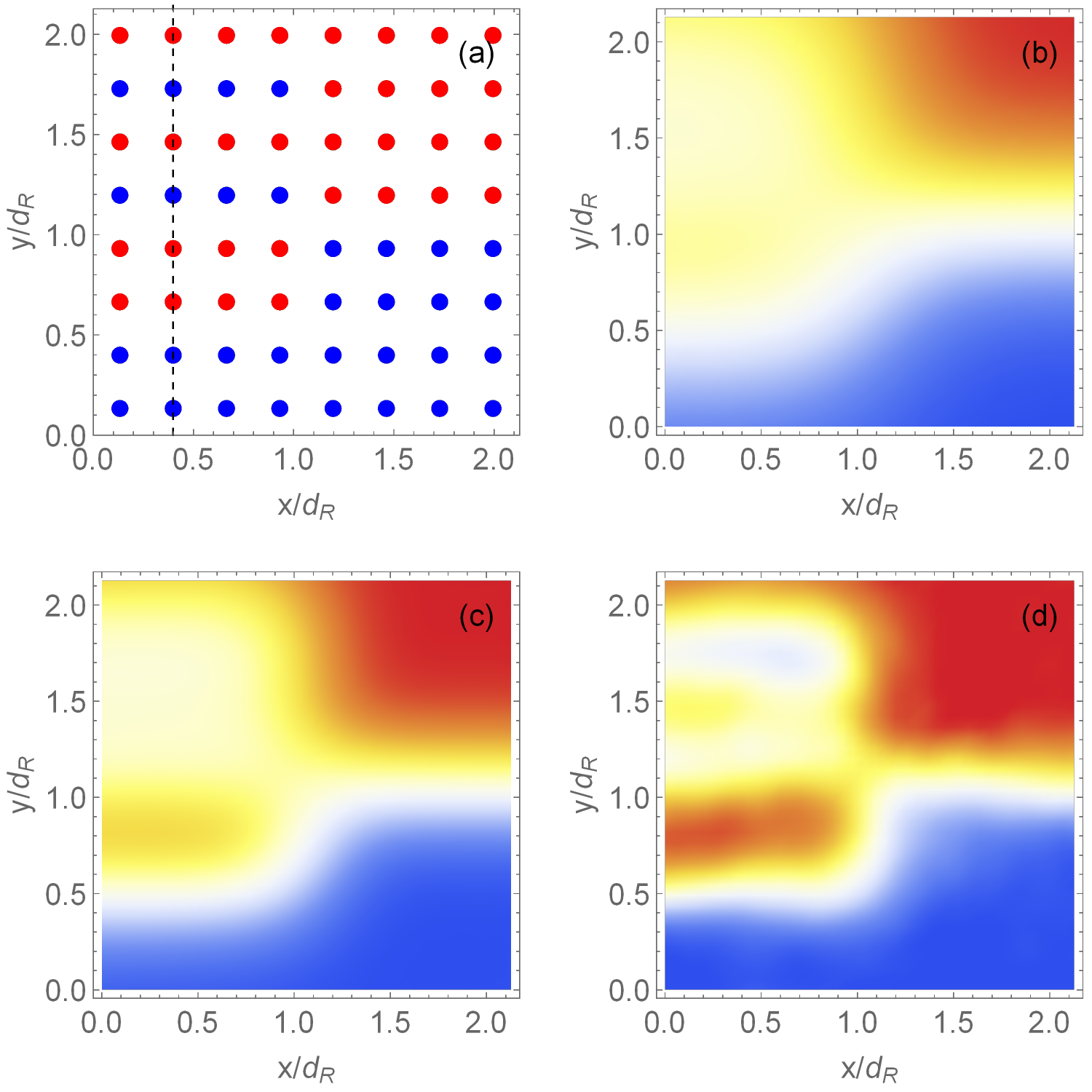}\\    
    \includegraphics[width=0.5\linewidth]{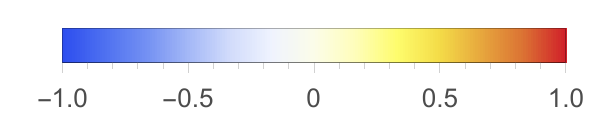}
    \caption{Modeling results for sub-Rayleigh imaging with sensing. (a) The object is modeled as a grid of emitters with the step $d = 0.266 d_\text{R}$. The emitters are assigned the parameter values $\theta = +1$ (red dots) and $\theta = -1$ (blue dots). The vertical dashed line shows the cross-section used for extracting the profiles in Fig.~\ref{fig:cuts}. (b-d) The reconstructed distribution $\bar \theta^{(n)}(\mathbf{r})$ of the sensed parameter for $n = 1$ (b), 2 (c), 4 (d). The color map is the same for all plots (blue and red colors correspond to $\bar \theta = -1$ and $\bar \theta = +1$ respectively). The spatial coordinates $x$ and $y$ are normalized by the Rayleigh limit $d_\text{R}$.}
    \label{fig:maps-blurred}
\end{figure}

\begin{figure}[tp]
    \centering
    \includegraphics[width=\linewidth]{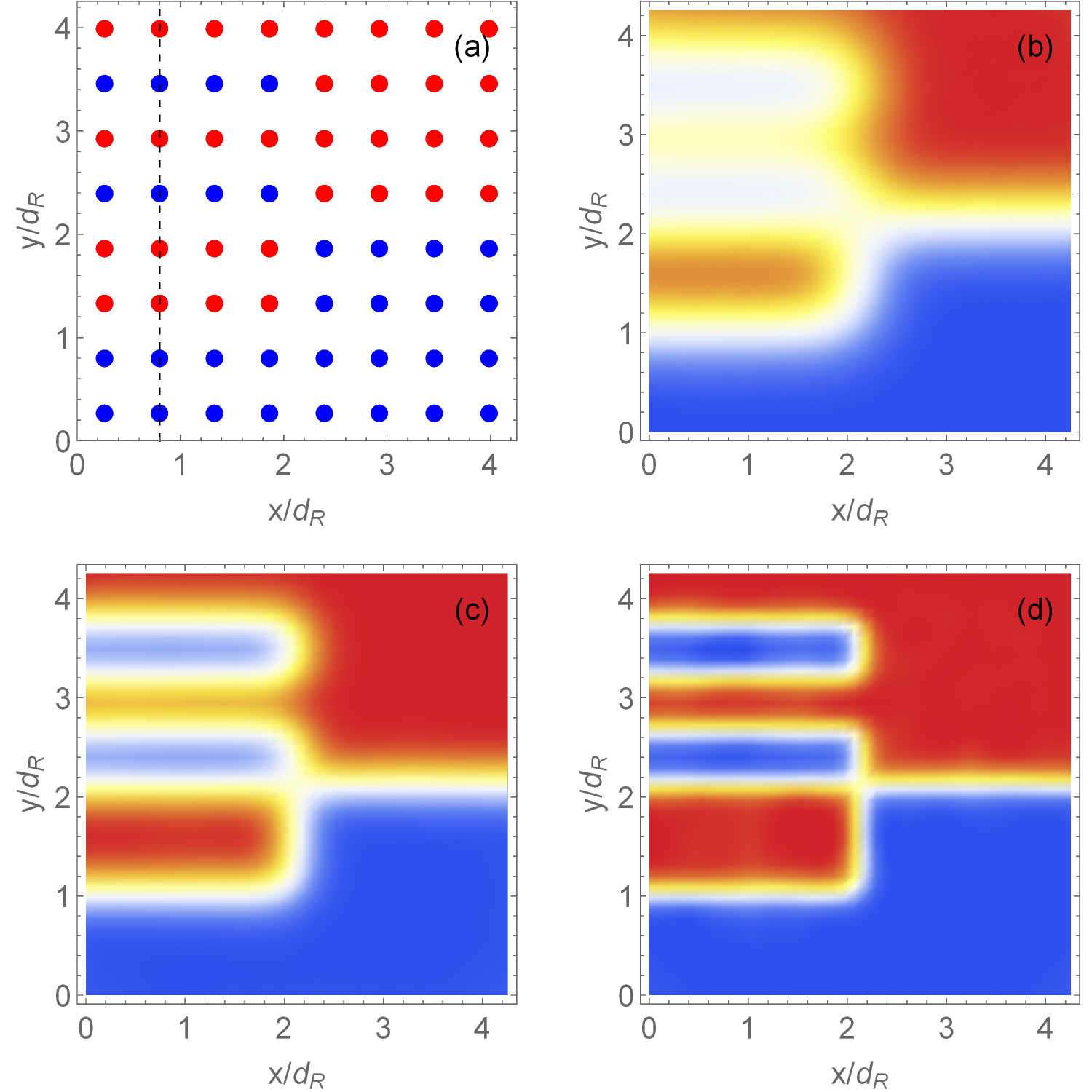}
    \caption{Modeling results for relatively well-resolved imaging with sensing. The step size of emitters' grid is $d = 0.532 d_\text{R}$. Other parameters, notations, and color legend are the same as in Fig.~\ref{fig:maps-blurred}.}
    \label{fig:maps-resolved}
\end{figure}

\begin{figure}[tp]
    \centering
    \includegraphics[width=\linewidth]{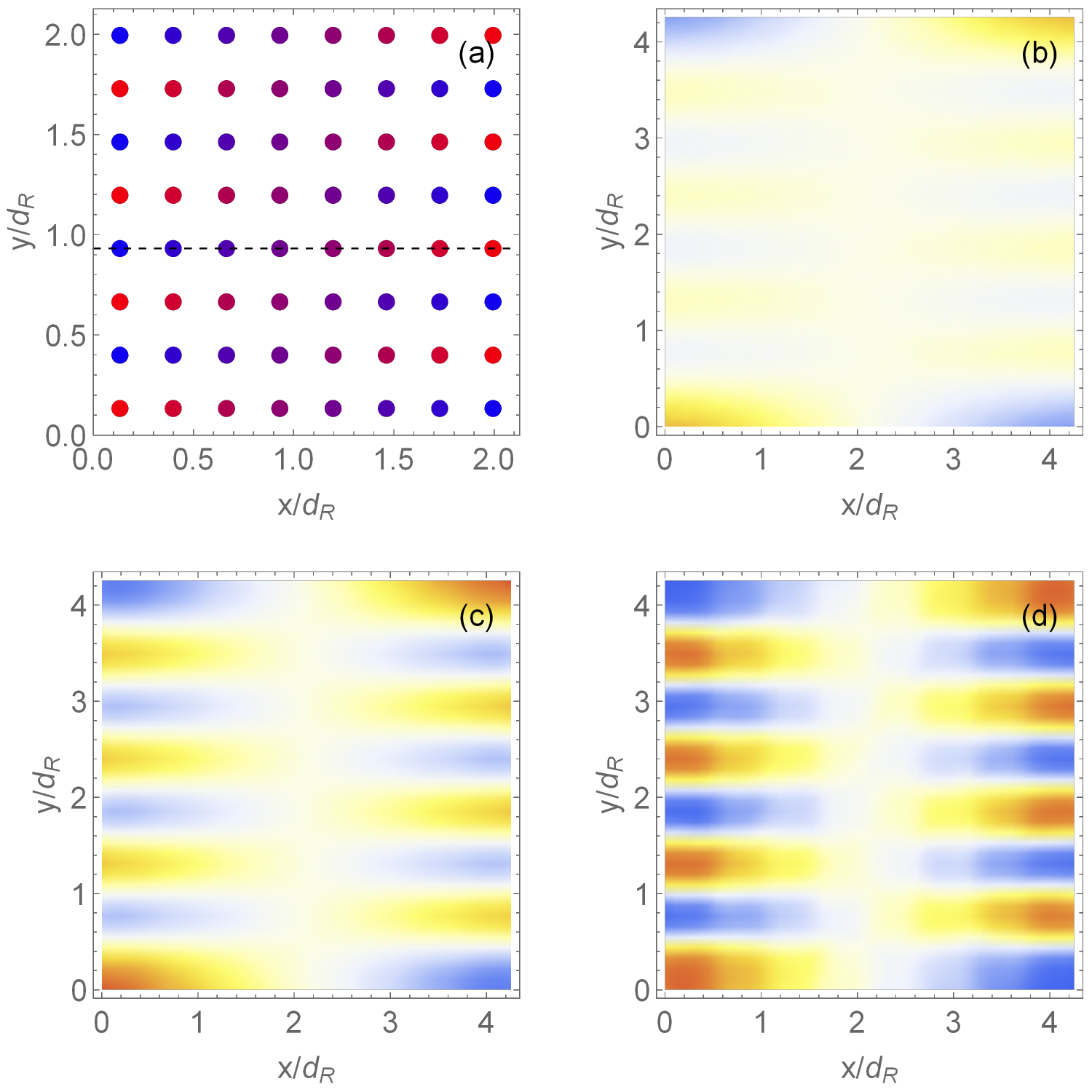}
    \caption{Modeling results for a sample with linear change of the sensed parameter $\theta$ along $x$ direction from $\theta = - 1$ (blue) to $\theta = + 1$ (red) or vice versa. The step size of emitters' grid is $d = 0.532 d_\text{R}$. Other parameters and notations are the same as in Fig.~\ref{fig:maps-blurred}.}
    \label{fig:maps-linear-resolved}
\end{figure}

The {\color{black}first} modeled object includes equidistant fluorescent probes in $8\times 8$ grid with the step $d = 0.266 d_\text{R}$ (Fig.~\ref{fig:maps-blurred}) or $0.532 d_R$ (Figs.~\ref{fig:maps-resolved} and \ref{fig:maps-linear-resolved}), where $d_\text{R}$ is the Rayleigh resolution limit of the microscope for classical incoherent-light imaging. {\color{black}The mean photon numbers correspond to the high-brightness regime. We take $\langle N \rangle = 1 \times 10^4$ photons per emitter per frame for Figs.~\ref{fig:maps-resolved} and \ref{fig:maps-linear-resolved}. For the case of large blurring (Fig.~\ref{fig:maps-blurred}), the SNR reduction caused by the increase of the emitters number in the PSF spot (see Subsection~\ref{subsec:SOFI limitations}) is compensated by higher brightness: we assume $\langle N \rangle = 4\times 10^4$ photons per emitter per frame.}

For the model samples, presented in Figs.~\ref{fig:maps-blurred}(a) and \ref{fig:maps-resolved}(a), each emitter is assigned the value $\theta_k = \pm 1$ of the sensed parameter $\theta$. The modeled spatial distribution of the parameter $\theta$ has been chosen so that sensing of features with different spatial scale can be analyzed (the sample resembles a resolution target in imaging). The sample in Fig.~\ref{fig:maps-linear-resolved}(a) is intended for demonstration of continuous-parameter sensing: the parameter $\theta$ changes linearly along $x$ direction from $\theta = - 1$ for the left border to $\theta = + 1$ for the right border or vice versa (the direction is alternating for rows of emitters). 

\begin{figure}[tp]
    \centering
    \includegraphics[width=0.8\linewidth]{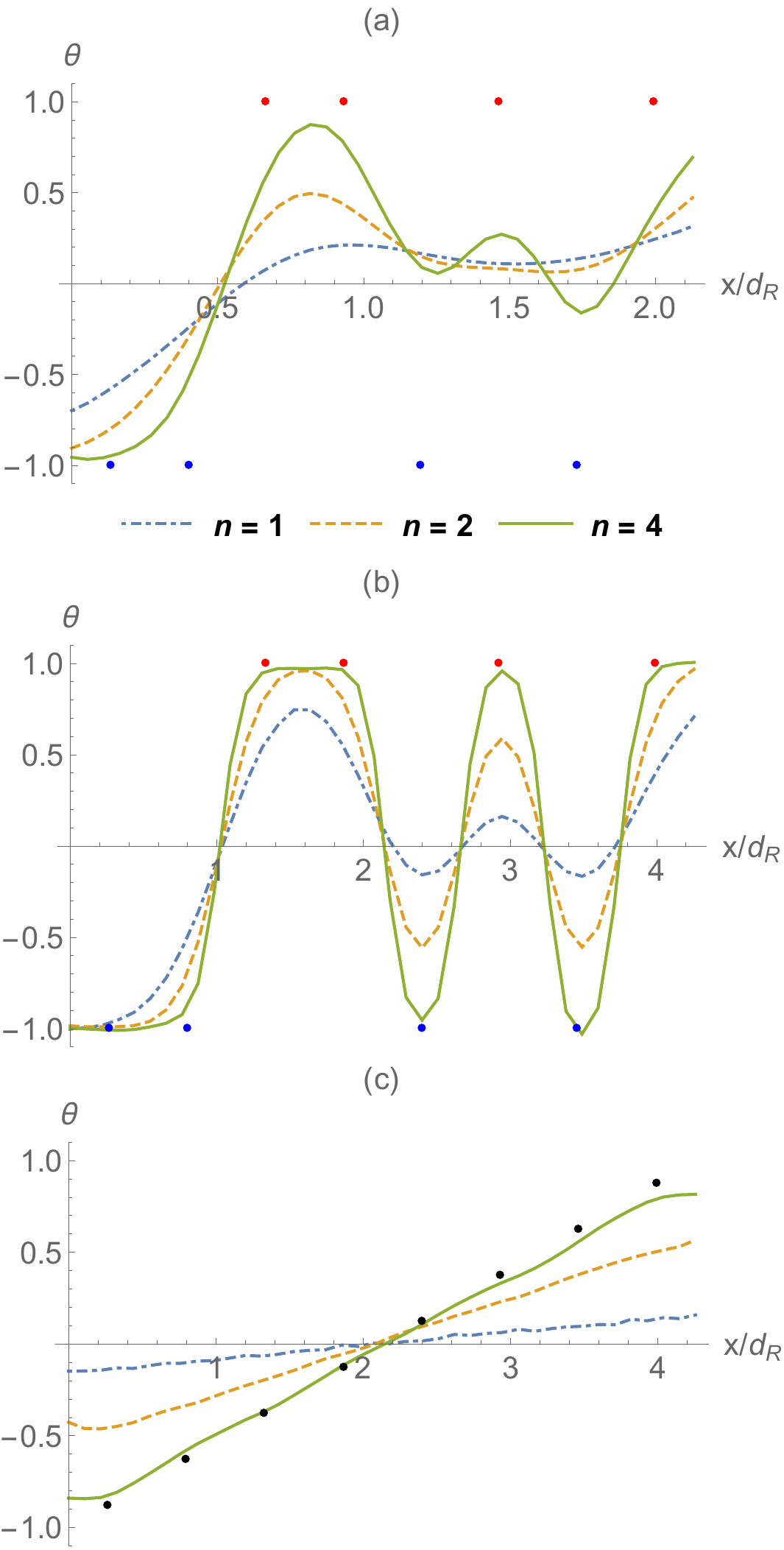}
    \caption{Inferred parameter distributions for the vertical cross-section of the maps, shown in Figs.~\ref{fig:maps-blurred} and \ref{fig:maps-resolved} --- panels (a) and (b) respectively, and the horizontal cross-section of the map in Fig.~\ref{fig:maps-linear-resolved} --- panel (c). Dot-dashed, dashed, and solid lines correspond to $n = 1$, 2, and 4 respectively. Dots indicate the true values of the parameter $\theta_k$ for the emitters along the considered sample cross-section.}
    \label{fig:cuts}
\end{figure}

Figs.~\ref{fig:maps-blurred}, \ref{fig:maps-resolved}, and \ref{fig:maps-linear-resolved} show the modeling results for the cases of sub-Rayleigh and relatively well-resolved imaging with sensing. The SOFI images are constructed according to the approach, described in Appendix~\ref{app:cross-cumulants} with $\sigma = 5$ (Fig.~\ref{fig:maps-blurred}) and $\sigma = 3$ (Figs.~\ref{fig:maps-resolved} and \ref{fig:maps-linear-resolved}). One can see that, as expected, the spatial resolution of the parameter inference increases for higher orders of SOFI cumulant images. To illustrate the effect, profiles corresponding to cross-sections of the maps are shown in Fig.~\ref{fig:cuts}. For sub-Rayleigh imaging regime (Fig.~\ref{fig:cuts}(a)), the fine features of the parameter distribution are not visible in the results for the 1st and the 2nd order cumulants, while separate minima and maxima are already present in the profile corresponding to the 4th order cumulant (i.e., the features are resolved according to Rayleigh's criterion). For the case shown in Fig.~\ref{fig:cuts}(b), the features of the parameter distribution are successfully resolved for all the considered cumulant orders. However, the accuracy of the parameter inference is much higher for the 4th order of SOFI (Table~\ref{tab:MSE}).

The results shown in Fig.~\ref{fig:maps-linear-resolved} and \ref{fig:cuts}(c) illustrate the main capability of our approach, clearly distinguishing it from color SOFI: accurate reconstruction of linear spatial distribution of the parameter $\theta$ (not just discrimination between the two extreme values $\theta = -1$ and $+1$). As one should expect, the 4th order cumulants yield more accurate results relatively to the 2nd and the 1st order.

\color{black}

Performance of the constructed cumulant-based estimator can be quantified by the mean squared error (MSE) of inferring the parameter values $\theta_k$ at the probe positions $\mathbf{r}_k$, defined as $\operatorname{MSE}(\bar \theta) = \langle(\bar\theta(\mathbf{r}_k) - \theta_k)^2 \rangle$. The results are summarized in Table~\ref{tab:MSE}. Generally, the performance represents a tradeoff between two target characteristics: SNR and resolution. Higher-order cumulants potentially yield better resolution, which, however, can be compromised by worth SNR. The interplay of the two effects and optimization of the cumulant order in terms of Fisher information has already been discussed in several papers \cite{vlasenko2020optimal,kurdzialek2021super,mikhalychev2025construction} devoted to SOFI (without sensing capabilities). Combination of cross-cumulants with autocumulants for improvement of SNR can make the higher-order cumulant images less noisy and shift the tradeoff (see also Ref.~\cite{peshko2025up}). For the considered model examples with high brightness of emitters, the resolution enhancement effect is prevailing, and the 4th order cumulant outperforms the 1st and the 2nd order ones.

\begin{table}[ht]
\color{black}
\caption{Mean square error $\operatorname{MSE}(\bar \theta)$ of the parameter $\theta$ estimation for the 1st, 2nd, and 4th order cumulants}
\label{tab:MSE}
\begin{tabular}{|p{0.47\linewidth}|c|c|c|}
        \hline
        Model & 1st order & 2nd order & 4th order\\ \hline
        Sub-Rayleigh imaging of 2D grid, Fig.~\ref{fig:maps-blurred} & 0.50 & 0.39 & 0.27 \\  \hline
        Well-resolved imaging of 2D grid, Fig.~\ref{fig:maps-resolved} & 0.24 & 0.065 & 0.0027 \\  \hline
        Well-resolved imaging of 2D grid, continuous distribution, Fig.~\ref{fig:maps-linear-resolved} & 0.20 & 0.053 & 0.0019 \\  \hline
        Low-brightness imaging of 2D grid, Fig.~\ref{fig:maps-low-brightness} & 0.24 & 0.089 & -- \\ \hline
        1D filaments, Fig.~\ref{fig:maps-filaments} & 0.49 & 0.39 & 0.29 \\  \hline
    \end{tabular}
\end{table}

\subsection{Low-brightness regime}
\label{subsec:low-brightness}

\begin{figure}[tp]
    \color{black}
    \centering
    \includegraphics[width=\linewidth]{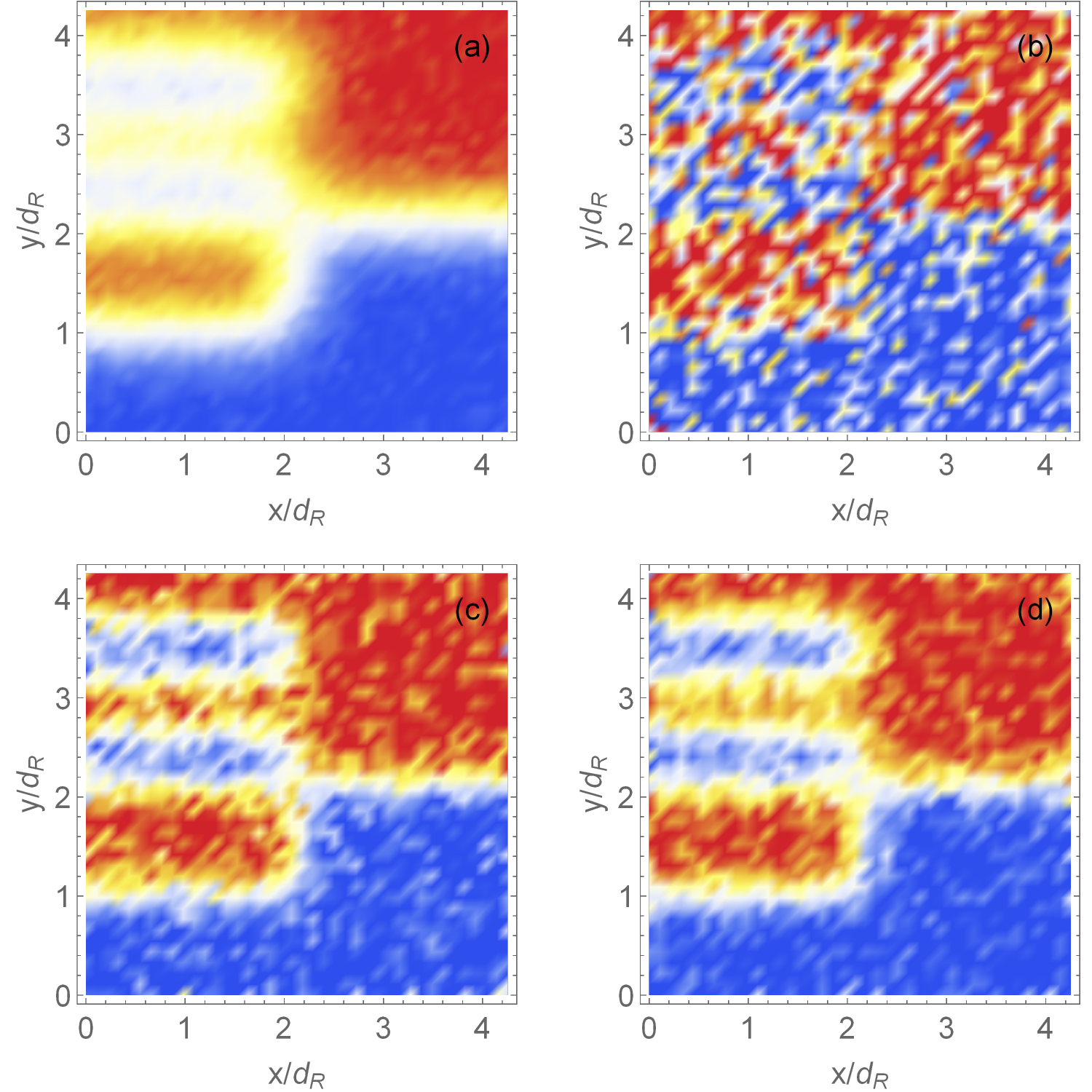}
    \caption{Modeling results for the object, shown in Fig.~\ref{fig:maps-resolved} in low-brightness regime. The reconstructed distribution $\bar \theta^{(n)}(\mathbf{r})$ of the sensed parameter is shown for $n = 1$ (a) and 2 (b-d). The 2nd order cumulant estimator is based on QSIPS with autocumulants (b) and cross-cumulants (Appendix~\ref{app:cross-cumulants} and Ref.~\cite{peshko2025up}) for $\sigma = 3$ (c) and 7 (d). The step size of emitters' grid is $d = 0.532 d_\text{R}$. Other parameters and notations are the same as in Fig.~\ref{fig:maps-blurred}.}
    \label{fig:maps-low-brightness}
\end{figure}

Fig.~\ref{fig:maps-low-brightness} shows the modeling results for 2-dimensional grid of emitters in the low-brightness regime. The configuration is the same as the one presented in Fig.~\ref{fig:maps-resolved}, but with 2500 times smaller intensity of the fluorescent signal ($\langle N \rangle = 4$ photons per emitter per frame) and 100 times larger number of frames. Since higher-order cumulants tend to suffer from the shot noise in the low-light regime \cite{picariello2025imaging,peshko2025up}, we limit the analysis by the 1st (Fig.~\ref{fig:maps-low-brightness}(a)) and the 2nd order (Fig.~\ref{fig:maps-low-brightness}(b-d)) QSIPS cumulants. Comparison of the parameter inference results for the auto-cumulant based estimator (Fig.~\ref{fig:maps-low-brightness}(b)) with the combined usage of auto- and cross-cumulants discussed in Appendix~\ref{app:cross-cumulants} (Fig.~\ref{fig:maps-low-brightness}(c,d)) clearly indicates improvement of the resulting SNR when the cross-cumulants are taken into account.

The presented results demonstrate that the developed approach preserves its efficiency in the low-light regime: construction of the parameter estimator on the base of the 2nd order cumulant improves the spatial resolution (Fig.~\ref{fig:maps-low-brightness}) and yields 2.7 times smaller MSE relatively to the intensity-based estimator (Table~\ref{tab:MSE}).

\subsection{1-dimensional filament-like structure}
\label{subsec:1D structure}

\begin{figure}[tp]
    \color{black}
    \centering
    \includegraphics[width=\linewidth]{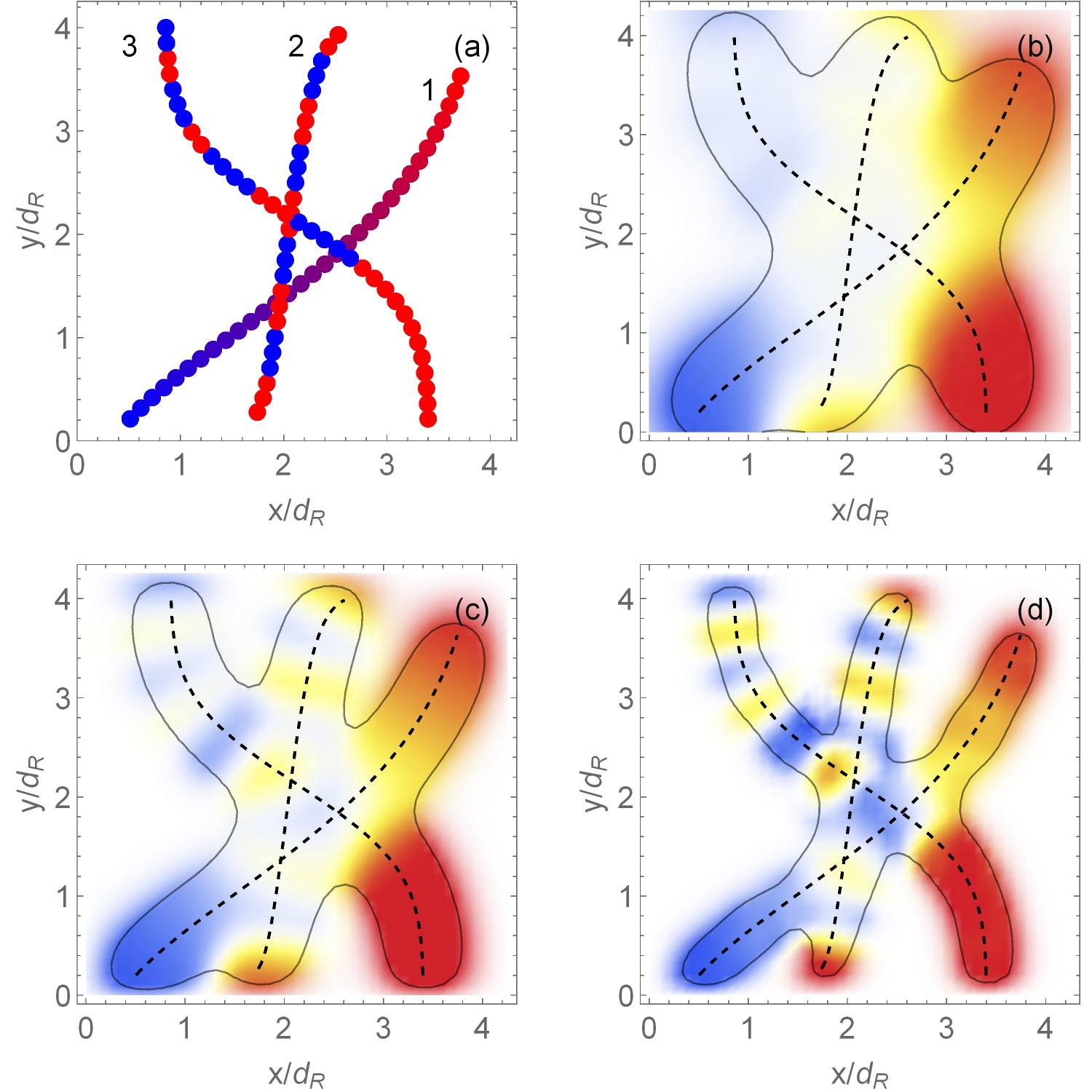}
    \caption{Modeling results for 1-dimensional arrangement of emitters. (a) The object is modeled as 3 filaments formed by equidistant emitters with the step $d = 0.15 d_\text{R}$. (b-d) The reconstructed distribution $\bar \theta^{(n)}(\mathbf{r})$ of the sensed parameter for $n = 1$ (b), 2 (c), 4 (d).  Other parameters and notations are the same as in Fig.~\ref{fig:maps-blurred}. Dashed lines show the positions of the modeled filaments for visual guidance. Thin solid lines indicate boundary of the region, where the signal is strong enough to provide reliable estimation of the sensed parameter $\theta$. For clarity of visualization, coloring of the maps is suppressed outside the region with the reliable parameter estimation.}
    \label{fig:maps-filaments}
\end{figure}

Discussion in Subsection~\ref{subsec:SOFI limitations} suggests that application of the SOFI-based sensing to 1-dimensional structures can yield reasonable SNR level even for emitters separated by essentially sub-Rayleigh distances $d \ll d_\text{R}$. Fig.~\ref{fig:maps-filaments}(a) shows a 3-filament model object with the distance between adjacent emitters along each filament equal to $d = 0.15 d_\text{R}$. The filaments are assigned three characteristic model distributions of the parameter $\theta$: continuous linear variation from $\theta = -1$ to $\theta = +1$ (filament 1 in Fig.~\ref{fig:maps-filaments}(a)), segments of equal sub-Rayleigh length $3d = 0.45 d_\text{R}$ with alternating parameter values $\theta = \pm 1$ (filament 2), and segments of unequal lengths with alternating parameter values (filament 3). High-brightness regime ($\langle N\rangle = 1 \times 10^4$ photons per emitter per frame, $1.5 \times 10^4$ frames) is assumed. 

The results presented in Fig.~\ref{fig:maps-filaments}(b-d) demonstrate successful resolution of the fine features by the estimator based on the 4th order QSIPS cumulant, while the intensity-based analysis fails to recover the structure of the sensed parameter distribution. The advantages of the developed cumulant-based approach are also confirmed by the behavior of MSE of the parameter inference, which takes the lowest value of the 4th order cumulant (Table~\ref{tab:MSE}).

\color{black}

\section{Conclusions}
\label{sec:conclusions}

We proposed a simple and efficient technique capable of combining sensing with super-resolution imaging, based on non-trivial temporal statistics typical for blinking fluorescent nanoemitters. The approach is universal since the dependence of the fluorophores response on the characteristic to be reconstructed can take different form: changes of the spectrum, fluorescence lifetime, etc. It is shown that splitting the fluorescence signal into just two detection channels at each spatial detection point is sufficient for reliable reconstruction of the local characteristic distribution.

\color{black}

When sensing is combined with SOFI, the approach inherits its limitations and operation conditions, such as the constraints on brightness and density of emitters. The former constraint can be released by combining sensing with QSIPS, which represents extension of SOFI to low-light (quantum) regime. It is worth noting that QSIPS does not require emitters to be blinking in the traditional sense (stochastically switching between bright and dark states): super- or sub-Poissonian statistics of emission is sufficient for achieving super-resolution \cite{picariello2025imaging,peshko2025up}. That capability opens route for applying the developed approach to up-converting nanoparticles or non-blinking color centers.

\color{black}

The conducted numerical simulations proof soundness and efficiency of the approach and can be used as a guidance for practical applications with, for example, living cells and structured fluorescent nanodiamonds \cite{quarshie2024nano} or blinking graphene quantum dots ensembles \cite{belko2022hysteresis}. In agreement with the theoretical predictions, usage of cumulants instead of emission intensity enhances sharpness of the reconstructed parameter distribution, enables analysis of smaller features of the sample, and ensures higher accuracy of the estimation.

Proposed methodology will be useful for local measurements of living cells' physical parameters, such as temperature, pH, electric field gradient, presence of reactive oxygen species, i.e. the characteristics which could be recognized as spectral changes of appropriate fluorophore  delivered targetly to the specific cell or even organelle. In its turn, this advanced local sensing tool could give an important insights into fundamental life science, which still lacks instruments capable of monitoring the cellular balance with nanometer spatial resolution to guarantee measurements consistency for that dynamic system.  

We believe that the proposed approach can pave the way for further research, including the following possible directions: 
\begin{itemize}
    \item \textit{Detailed analysis and optimization of the combined sensing and imaging approach in terms of classical and quantum Fisher information.} {\color{black}While being relatively simple for quantum imaging with Poissonian statistics of signal \cite{mikhalychev2019efficiently,mikhalychev2021lost,mikhalychev2025efficient}, construction of multiparametric Fisher information matrix becomes extremely difficult in SOFI since blinking leads to correlated non-Poissonian noise in cumulant images and may require high-performance computing \cite{mikhalychev2025construction}.}  
    \item \textit{Reconstruction of the sample characteristic distribution by combined usage of several orders of cumulants simultaneously.} {\color{black}The procedure is relatively straightforward and follows the ideas of Subsection~\ref{subsec:noisy}: the parameter estimates from the available cumulant orders should be weighted according to their inaccuracies (see also Refs.~\cite{kurdzialek2021super,mikhalychev2025construction}). However, reliable comparison of inaccuracies for different cumulant orders represents a separate complicated task, which can be implemented by construction of Fisher information matrix and usage of Cramer-Rao bound.}
    \item \textit{Extension of the approach to multimodal sensing, when several characteristics of the sample are reconstructed simultaneously.} {\color{black}To estimate $N$ parameters, one needs to acquire $(N + 1)$-channel signal. Selection of the optimal channels is similar to the procedure discussed in Appendix~\ref{app:optimal_estimator}, but requires replacement of scalar Fisher information by Fisher information matrix.}
    \item \textit{Implementation of Fourier spectrum re-weighting   for additional de-blurring and cumulant orders combination.} {\color{black} The procedure is similar to combining several orders of cumulants and requires estimation of reliability for components of Fourier spectrum.}
    \item \textit{Application of the ``sliding window'' approach \cite{mikhalychev2019efficiently} for more efficient processing of many-pixel images.} {\color{black}The explicit parameter estimator (\ref{eq:estimator_SOFI}) can be replaced by a set of optimization tasks, where the parameter values $\bar \theta^{(n)}(\mathbf{r})$ are found by local fitting of the 2-channel cumulant images $C_1^{(n)}(\mathbf{r})$ and $C_0^{(n)}(\mathbf{r})$. The technique can yield a sharper (more accurate) distribution of the sensed parameter, but requires substantial computational efforts.}
\end{itemize}

\begin{acknowledgements}
    This study was accomplished through the financial support of Horizon Europe MSCA FLORIN Project 101086142.
\end{acknowledgements} 

\appendix

\section{Locally optimal signal compression}
\label{app:optimal_estimator}

Here, we show that the information about the environment parameter $\theta$, contained in the normalized signal density profile $g(x) \equiv \tilde f(\mathbf{r}_0, t, x)$, defined by Eq.~(\ref{eq:normalized signal}), can be efficiently compressed into a ratio $Z$ of two scalar characteristics $Y_i$ (Eq.~(\ref{eq:Z definition})) containing the same information. We assume that the signal profile is measured at a discrete set of bins $\{ x \}$ and contains shot noise $\delta(x)$ satisfying Poisson statistics (i.e., we analyze shot noise for a fixed blinking realization since the blinking influences the in-frame signal profile as a whole):
\begin{equation}
    g(x) \mapsto g_\text{noisy}(x) = g(x) + \delta(x),
\end{equation}
where $\operatorname{E}[\delta(x)] = 0$, $\operatorname{E}[\delta(x)\delta(x')] \approx g(x) \delta_{xx'}$. The high-SNR regime is considered, i.e., $|\delta(x)| \ll |g(x)|$. 

The Fisher information \cite{fisher1925theory} of the normalized density profile $g(x)$ relatively to the parameter $\theta$ can be calculated as
\begin{equation}
    F_\text{full}(\theta) = \sum_x \frac{\left[g'(x)\right]^2}{g(x)} - \frac{1}{\sum_x g(x)} \Bigl( \sum_x g'(x) \Bigr)^2,
\end{equation}
where the second term describes ``information loss'' caused by the normalization \cite{mikhalychev2021fisher} and $g'(x) = dg(x) / d\theta$.

Substitution of $g_\text{noisy}(x)$ into Eq.~(\ref{eq:Z definition}) and replacement of the integrals by sums yields
\begin{multline}
    Z \mapsto Z_\text{noisy} \\ \approx Z \left(1 + \frac{\sum_x \delta(x) Q_1(x)}{\sum_x g(x) Q_1(x)}  - \frac{\sum_x \delta(x) Q_0(x)}{\sum_x g(x) Q_0(x)} \right).
\end{multline}
In the considered high-SNR regime, $Z_\text{noisy}$ represents a normally distributed random variable with the expectation value $\operatorname{E}(Z_\text{noisy}) = Z$ and the variance
\begin{equation}
    \label{eq:app:Var Z}
    \operatorname{Var}(Z_\text{noisy}) = \frac{1}{Y_0^2} \sum_x g(x) \bigl[Q_1(x) - Q_0(x) Z\bigr]^2.
\end{equation}
Therefore, Fisher information for the signal $Z_\text{noisy}$ relatively to the sensed parameter $\theta$ equals (see, e.g., Ref.~\cite{mikhalychev2021fisher})
\begin{equation}
    \label{eq:app:F_Z}
    F_Z(\theta) = \frac{1}{\operatorname{Var}(Z_\text{noisy})}\left( \frac{dZ}{d\theta} \right)^2.
\end{equation}

According to general properties of Fisher information \cite{Lehmann-Casella,Cohen68}, the compressed signal $Z_\text{noisy}$ cannot contain more information than the initial normalized profile $g_\text{noisy}(x)$ for any choice of the weights $Q_0(x)$ and $Q_1(x)$:
\begin{equation}
    F_{Z} (\theta) \le  F_\text{full}(\theta).
\end{equation}
On the other hand, by a specific choice of those weights, one can achieve equality of the two quantities. One can define
\begin{equation}
    \label{eq:app:optimal_q}
    Q_0(x) = \frac{q_0(x)}{\sum_{x'} q_0(x') g(x')}, \; Q_1(x) = \frac{q_1(x)}{\sum_{x'} q_1(x') g'(x')},
\end{equation}
where
\begin{equation}
    q_0(x) = g(x) - g'(x)\frac{\sum_{x'} g(x') g'(x')}{\sum_{x'} \left[g'(x')\right]^2}
\end{equation}
and
\begin{equation}
    q_1(x) = \frac{g'(x)}{g(x)} - \frac{\sum_{x'} g'(x')}{\sum_{x'} g(x')}.
\end{equation}
That specific choice of the weights ensures that
\begin{equation}
    \label{eq:app:Ys optimal}
    Y_0 = 1, \quad Y_1 = 0, \quad \frac{dY_0}{d\theta} = 0, \quad \frac{dY_1}{d\theta} = 1
\end{equation}
and 
\begin{equation}
    \label{eq:app:Z optimal}
    Z =0, \quad  \frac{dZ}{d\theta} = 1
\end{equation}
at the considered value of the parameter $\theta$ (the one used for construction of the weights).
Substituting Eqs.~(\ref{eq:app:Ys optimal}) and (\ref{eq:app:Z optimal}) into Eqs.~(\ref{eq:app:Var Z}) and (\ref{eq:app:F_Z}), we get the results $\operatorname{Var}(Z_\text{noisy}) = 1/ F_\text{full}(\theta)$ and
\begin{equation}
    \label{eq:app:equality of FIMs}
    F_{Z} (\theta) = F_\text{full}(\theta).
\end{equation}

The choice of the weights is \textit{locally} optimal. Generally, for values of $\theta$, other than the specific one used for the weights construction, the equality (\ref{eq:app:equality of FIMs}) may be violated. It is worth noting that in the high-SNR limit and in the vicinity of the considered parameter value $\theta$ the estimator 
\begin{equation}
    \bar \theta = \zeta^{-1}(Z) \approx \theta + Z
\end{equation}
with the weights described by Eq.~(\ref{eq:app:optimal_q}) coincidence with the locally unbiased estimator, saturating Cramer-Rao bound \cite{demkowicz2020multi}.

\section{General properties of cumulant-based estimator for noiseless data}
\label{app:cumulant-based_estimator}

To be physically reasonable, the estimator $\bar \theta^{(n)}(\mathbf{r})$, defined by Eq.~(\ref{eq:estimator_SOFI}) should satisfy several general requirements when applied to noiseless data. The requirements are provided by Eqs.~(\ref{eq:estimator uniform})--(\ref{eq:estimator linear approx.}). Here, we proof those statements.

1. For a spatially uniform distribution $\theta(\mathbf{r}) = \theta_0$, one has $Y_j(\theta_k) = Y_j(\theta_0)$ for all emitter indices $k$. Therefore, according to Eq.~(\ref{eq:cumulant_for_window}),
\begin{equation}
    \frac{C_1^{(n)}(\mathbf{r})}{C_0^{(n)}(\mathbf{r})} = \frac{Y_1^n(\theta_0)}{Y_0^n(\theta_0)} = \zeta^n(\theta_0)\text{ for all $\mathbf{r}$},
\end{equation}
and
\begin{equation}
    Z_n(\mathbf{r}) = \zeta(\theta_0) \quad \Rightarrow \quad \bar \theta^{(n)}(\mathbf{r}) = \theta_0.
\end{equation}

2. If the parameter values $\theta_k$ in the vicinity of the emitters are bounded as $\theta_k \in [\theta_\text{min}, \theta_\text{max}]$, the monotonicity assumption implies that the quantities $\zeta(\theta_k)$ are also bounded as $\zeta(\theta_k) \in [\zeta_\text{min}, \zeta_\text{max}]$, where $\zeta_\text{min} = \min(\zeta(\theta_\text{min}), \zeta(\theta_\text{max}))$ and $\zeta_\text{max} = \max(\zeta(\theta_\text{min}), \zeta(\theta_\text{max}))$. Under the assumption that $\zeta_\text{min} \ge 0$ and $U(\mathbf{r} - \mathbf{r}_k) \ge 0$, for each emitter, the following relation holds:
\begin{equation}
    Y_1^n(\theta_k) \in Y_0^n(\theta_k) \times [\zeta_\text{min}^n, \zeta_\text{max}^n].
\end{equation}
Therefore, for any $\mathbf{r}$,
\begin{equation}
    C_1^{(n)}(\theta_k) \in C_0^{(n)}(\theta_k) \times [\zeta_\text{min}^n, \zeta_\text{max}^n].
\end{equation}
Finally, $Z_n(\mathbf{r}) \in [\zeta_\text{min}, \zeta_\text{max}]$ and, therefore, $\bar \theta^{(n)}(\mathbf{r}) \in [\theta_\text{min}, \theta_\text{max}]$.

3. If the variations $\delta \theta_k = \theta_k - \theta_0$ are small, one can use the linear approximation and get 
\begin{equation}
    \label{eq:app:Yjn linear}
    Y_j^n(\theta_k) \approx Y_j^n(\theta_0) + n Y_j^{n-1}(\theta_0) Y_j'(\theta_0) \delta \theta_k,
\end{equation}
where $Y_j'(\theta) = \int \frac{\partial \varepsilon(x, \theta)}{\partial \theta} Q_j(x) dx$.
Substitution of Eq.~(\ref{eq:app:Yjn linear}) into Eq.~(\ref{eq:cumulant_for_window}) yields
\begin{multline}
    C_j^{(n)}(\mathbf{r}) \approx Y_j^n(\theta_0) \sum_k U^n(\mathbf{r} - \mathbf{r}_k) \varepsilon_k^n  C_t^{(n)}[s_k(t)] \\ + n Y_j^{n-1}(\theta_0) Y_j'(\theta_0) \sum_k U^n(\mathbf{r} - \mathbf{r}_k) \varepsilon_k^n C_t^{(n)}[s_k(t)] \delta \theta_k 
\end{multline}
and 
\begin{multline}
    Z_n(\mathbf{r}) \approx \zeta(\theta_0) \Biggl[1 + \left(\frac{Y_1'(\theta_0)}{Y_1(\theta_0)} - \frac{Y_0'(\theta_0)}{Y_0(\theta_0)}\right) \\ \times \frac{\sum_k U^n(\mathbf{r} - \mathbf{r}_k) \varepsilon_k^n C_t^{(n)}[s_k(t)] \delta \theta_k}{\sum_k U^n(\mathbf{r} - \mathbf{r}_k) \varepsilon_k^n C_t^{(n)}[s_k(t)]}\Biggr]
\end{multline}

Since
\begin{equation}
    \zeta(\theta_0 + \delta \theta) \approx \zeta(\theta_0) \left[1 + \left(\frac{Y_1'(\theta_0)}{Y_1(\theta_0)} - \frac{Y_0'(\theta_0)}{Y_0(\theta_0)}\right) \delta \theta \right],
\end{equation}
one can easily find the solution of the equation $\zeta(\theta) = Z_n(\mathbf{r})$ in the linear approximation and conclude that
\begin{multline}
    \bar \theta^{(n)}(\mathbf{r}) = \theta_0 + \frac{\sum_k U^n(\mathbf{r} - \mathbf{r}_k) \varepsilon_k^n C_t^{(n)}[s_k(t)] \delta \theta_k}{\sum_k U^n(\mathbf{r} - \mathbf{r}_k) \varepsilon_k^n C_t^{(n)}[s_k(t)]} \\ = \frac{\sum_k U^n(\mathbf{r} - \mathbf{r}_k) \varepsilon_k^n C_t^{(n)}[s_k(t)] \theta_k}{\sum_k U^n(\mathbf{r} - \mathbf{r}_k) \varepsilon_k^n C_t^{(n)}[s_k(t)]}.
\end{multline}

\section{Expressions for regularized parameter estimation}
\label{app:regularized_expressions}
For parameter values, defined at a regular grid $\mathbf{r}_{ij}$, the derivatives in Eq.~(\ref{eq:regularizing_p0}) can be approximated as finite differences. Then, the expression for log-probability takes the form
\begin{multline}
    \label{eq:app:log_p}
    - \log p\left[\theta(\mathbf{r}_{ij}) | \bar \theta^{(n)}(\mathbf{r}_{ij})\right] \propto \sum_{i,j} \frac{[\bar \theta^{(n)}(\mathbf{r}_{ij}) - \theta(\mathbf{r}_{ij})]^2}{2 \left[\Delta \bar \theta^{(n)}(\mathbf{r}_{ij})\right]^2}
    \\ + \sum_{i,j} \frac{[2 \theta(\mathbf{r}_{ij}) - \theta(\mathbf{r}_{i-1,j}) - \theta(\mathbf{r}_{i+1,j})]^2}{2 D^2}
    \\ + \sum_{i,j} \frac{[2 \theta(\mathbf{r}_{ij}) - \theta(\mathbf{r}_{i,j-1}) - \theta(\mathbf{r}_{i,j+1})]^2}{2 D^2}.
\end{multline}
The sums representing the second and the third terms do not include boundary values of $i$ and $j$ respectively. The parameter $D$ scales as $D = \text{const} \times \langle \Delta \bar \theta^{(n)}(\mathbf{r}_{ij}) \rangle w$, where $w$ is the PSF width expressed in the grid steps and representing the expected minimal size of the reconstructed distribution features, which are not artifacts yet. The proportionality coefficient can be calculated from detailed theoretical investigation of the noise models, but, in practice, it can be just chosen empirically.

Unconstrained minimization of the quantity, defined by the right-hand side of Eq.~(\ref{eq:app:log_p}), is trivial and consists in linear inversion according to the least-squares method. One of the approaches to obtaining physically reasonable constrained estimates $\theta_\text{min} \le \theta(\mathbf{r}_{ij}) \le \theta_\text{max}$ is to perform constrained quadratic optimization, which, however, becomes much more computationally hard for a large number of grid points $\mathbf{r}_{ij}$. A more efficient way to get reasonable results is to constrain the initial estimate $\bar \theta^{(n)}(\textbf{r}_{ij})$ to the range $[\theta_\text{min}, \theta_\text{max}]$ before applying the minimization of the expression (\ref{eq:app:log_p}). If all the values satisfy the constraint $\bar \theta^{(n)}(\textbf{r}_{ij}) \in [\theta_\text{min}, \theta_\text{max}]$, the minimizing set $\{\theta(\textbf{r}_{ij}\}$ will also belong to the physical range with certainty.

\section{SOFI with cross-cumulants}
\label{app:cross-cumulants}

The expression~(\ref{eq:SOFI_cross-cumulant}) for a the zero time lag cross-cumulant
\begin{multline}
    C_t^{(n)} [F(\mathbf{r}^{(1)}, t), \ldots, F(\mathbf{r}^{(n)}, t)] \\ \propto \sum_k U^n(\bar{\mathbf{r}} - \mathbf{r}_k) \varepsilon_k^n C_t^{(n)}[s_k(t)],
\end{multline}
coincides with Eq.~(\ref{eq:SOFI_cumulant}) for an autocumulant $C_t^{(n)}[F(\bar{\mathbf{r}},t)]$ up to a constant multiplier. Therefore, to improve SNR, one can replace the autocumulant signal $C_t^{(n)}(\mathbf{r})$ used in SOFI by a weighted combination of cross-cumulants:
\begin{multline}
    \label{eq:app:weighted cross-cumulants}
    C_t^{(n)}(\mathbf{r}) \mapsto \tilde C_t^{(n)}(\mathbf{r}) = \sum_{
    \begin{array}{c}
         \scriptstyle \mathbf{r}^{(1)},\ldots, \mathbf{r}^{(n)}:  \\
         \scriptstyle \mathbf{r}^{(1)} + \cdots + \mathbf{r}^{(n)} = n \mathbf{r} 
    \end{array}} w_{\mathbf{r}^{(1)},\ldots, \mathbf{r}^{(n)}} \\ \times C_t^{(n)} [F(\mathbf{r}^{(1)}, t), \ldots, F(\mathbf{r}^{(n)}, t)].
\end{multline}

During processing of the simulated data, we choose the weights as
\begin{equation}
    w_{\mathbf{r}^{(1)},\ldots, \mathbf{r}^{(n)}} \propto \exp\left(- \frac{\sum_j \left|\mathbf{r}^{(j)}\right|^2}{\sigma^2}\right)
\end{equation}
with $\sigma = 3 \delta$, where $\delta$ is the detection step (detector pixel size, recalculated to the object plane of the microscope). As a compromise between noise reduction and computational complexity, we additionally restrict the sum in Eq.~(\ref{eq:app:weighted cross-cumulants}) by the condition $\sum_j \left|\mathbf{r}^{(j)}\right|^2 \le 5 \sigma^2$.

The approach can also be extended to non-zero time lag cross-cumulants, but we do not consider such modification since the goal of simulations is illustration of the proposed sensing approach rather than achievement of the ultimate accuracy of SOFI results.

The source code of the described method implementation is available in Supplemental Material at [URL will be inserted by publisher].

\bibliography{references.bib}

\end{document}